\documentclass[preprint,12pt]{aastex}
\usepackage{natbib}
\usepackage{graphicx}
\usepackage{latexsym}

\begin{document}

\title{The Mid-Infrared Instrument for the James Webb Space Telescope, VI: The Medium Resolution 
Spectrometer}

\author{Martyn Wells\altaffilmark{1},  J.-W.  Pel\altaffilmark{2},  Alistair Glasse\altaffilmark{1}, G. S. Wright\altaffilmark{1}, Gabby  Aitink-Kroes\altaffilmark{3}, Ruym\'an Azzollini\altaffilmark{4,5}, Steven Beard\altaffilmark{1}, B. R. Brandl\altaffilmark{6}, 
Angus Gallie\altaffilmark{1}, V. C. Geers\altaffilmark{4}, A. M. Glauser\altaffilmark{7}, Peter Hastings\altaffilmark{1}, Th. Henning\altaffilmark{8}, Rieks Jager\altaffilmark{3}, K. Justtanont\altaffilmark{9},  Bob Kruizinga\altaffilmark{10},  Fred Lahuis\altaffilmark{6, 11}, David Lee\altaffilmark{1}, 
I. Martinez-Delgado\altaffilmark{5}, J. R. Mart\'inez-Galarza\altaffilmark{6, 12}, M. Meijers\altaffilmark{3}, Jane E. Morrison\altaffilmark{13}, Friedrich M\"uller\altaffilmark{8}, Thodori Nakos\altaffilmark{14}, Brian O'Sullivan\altaffilmark{15},  Ad Oudenhuysen\altaffilmark{3},  P. Parr-Burman\altaffilmark{1}, Evert Pauwels\altaffilmark{3,16}, 
R.-R. Rohloff\altaffilmark{8}, Eva Schmalzl\altaffilmark{6}, Jon Sykes\altaffilmark{17}, M. P. Thelen\altaffilmark{18}, E. F. van Dishoeck\altaffilmark{6}, Bart Vandenbussche\altaffilmark{19}, Lars B. Venema\altaffilmark{16}, Huib Visser\altaffilmark{10}, L. B. F. M. Waters\altaffilmark{11, 20},  \& David Wright\altaffilmark{21}
}

\altaffiltext{1}{UK Astronomy Technology Centre, Royal Observatory,  Blackford Hill Edinburgh, EH9 3HJ, Scotland, United Kingdom}
\altaffiltext{2}{Kapteyn Institute, University of Groningen, PO Box 800, 9700 Groningen, The Netherlands}
\altaffiltext{3}{NOVA Opt-IR group, PO Box 2, 7990 AA Dwingeloo, The Netherlands}
\altaffiltext{4}{Dublin Institute for Advanced Studies, School of Cosmic Physics, 31 Fitzwilliam Place, Dublin 2, Ireland}
\altaffiltext{5}{Centro de Astrobiolog\'ia (INTA-CSIC), Depto Astrof\'isica, Carretera de Ajalvir, km 4, 28850 Torrej\'on de Ardoz, Madrid, Spain}
\altaffiltext{6}{Leiden Observatory, Leiden University, PO Box 9513, 2300 RA, Leiden, The Netherlands.}
\altaffiltext{7}{ETH Zurich, Institute for Astronomy, Wolfgang-Pauli-Str. 27, CH-8093 Zurich, Switzerland}
\altaffiltext{8}{Max Planck Institute f\"ur Astronomy (MPIA), K\"onigstuhl 17, D-69117 Heidelberg, Germany}
\altaffiltext{9}{Chalmers University of Technology, Onsala Space Observatory, S-439 92 Onsala, Sweden}
\altaffiltext{10}{TNO Optics, P.O. Box 155, 2600 AD Delft, The Netherlands.}
\altaffiltext{11}{SRON Netherlands Institute for Space Research, Sorbonnelaan 2, 3584 CA Utrecht, The Netherlands}
\altaffiltext{12}{Harvard-Smithsonian Center for Astrophysics, 60 Garden Street, Cambridge, MA 02138, USA}
\altaffiltext{13}{Steward Observatory, 933 N. Cherry Ave, University of Arizona, Tucson, AZ 85721, USA}
\altaffiltext{14}{Sterrenkundig Observatorium UGent, Krijgslaan 281 S9, B-9000 Gent, Belgium}
\altaffiltext{15}{Airbus Defence and Space, Anchorage Road, Portsmouth, Hampshire, PO3 5PU}
\altaffiltext{16}{Pi Environments B.V., Groen van Prinstererweg 57, 2221 AC, Katwijk, The Netherlands}
\altaffiltext{17}{Department of Physics and Astronomy, Univ. of Leicester, University Road, Leicester, LE1 7RH, UK}
\altaffiltext{18}{Jet Propulsion Laboratory, California Institute of Technology, 4800 Oak Grove Dr. Pasadena, CA 91109, USA}
\altaffiltext{19}{Institute of Astronomy KU Leuven, Celestijnenlaan 200D,3001 Leuven, Belgium}
\altaffiltext{20}{Sterrenkundig Instituut Anton Pannekoek, University of Amsterdam, Science Park 904, 1098 Amsterdam, The Netherlands}
\altaffiltext{21}{Stinger Ghaffarian Technologies, Inc., Greenbelt, MD, USA}
\altaffiltext{}{}
\altaffiltext{}{}
\altaffiltext{}{}
\altaffiltext{}{}
\altaffiltext{}{}
\altaffiltext{}{}
\altaffiltext{}{}
\altaffiltext{}{}
\altaffiltext{}{}
\altaffiltext{}{}
\altaffiltext{}{}
\altaffiltext{}{}

\begin{abstract}
We describe the design and performance of the Medium 
Resolution Spectrometer (MRS) for the JWST-MIRI instrument. The MRS 
incorporates four coaxial spectral channels in a compact opto-mechanical 
layout that generates spectral images over fields of view up to 7.7 x 7.7 
arcseconds in extent and at spectral resolving powers ranging from 1,300 to 
3,700. Each channel includes an all-reflective integral field unit (IFU): an 
`image slicer' that reformats the input field for presentation to a grating 
spectrometer. Two 1024 x 1024 focal plane arrays record the output spectral 
images with an instantaneous spectral coverage of approximately one third of 
the full wavelength range of each channel. The full 5 to 28.5 $\mu $m spectrum is 
then obtained by making three exposures using gratings and pass-band-determining 
filters that are selected using just two three-position mechanisms. 
The expected on-orbit optical performance is presented, based on testing of 
the MIRI Flight Model and including spectral and spatial coverage and 
resolution. The point spread function of the reconstructed images is shown 
to be diffraction limited and the optical transmission is shown to be 
consistent with the design expectations.
\end{abstract}

\section{Introduction}

The rationale for, capabilities of, and 
scientific context of the mid-infrared instrument (MIRI) on JWST are described 
in Rieke et al. (2014; hereafter Paper I) and the overall 
optical, thermal, mechanical/structural electronic and control aspects of 
the design are summarized in Wright et al. (2014; hereafter Paper II). This paper describes in more detail 
the MIRI Medium Resolution Spectrometer (MRS), which is an Integral Field 
Unit (IFU) spectrometer that covers the wavelength range from 5 to 28.5 
$\mu $m at spectral resolving powers of a few thousand. The MRS consists of 4 
channels that have co-aligned fields of view, simultaneously observing the 
wavelength ranges listed in Table 1 with 
individually optimised IFUs, collimators and gratings. 

Section \ref{sec:optical} of this paper provides a description of 
the optical design, including the rationale for choosing the IFU concept and 
its impact on how observations are carried out. Section 
\ref{sec:measured} then describes the expected on-orbit optical 
performance of the MRS as measured during 
cryogenic testing, and with a description of the procedure used to construct 
calibrated spectral data cubes from the raw measured images. The impact of 
particular characteristics of the MRS, including spectral fringing and straylight 
are also discussed here.

\section{Optical Design}
\label{sec:optical}
\subsection{Rationale}
An IFU based design was preferred to a long-slit design for the MIRI 
Spectrometer for the following reasons.

Firstly, for point source observations, the need to centre the source in a 
narrow slit (via a peak-up procedure) is relaxed, simplifying and accelerating the target acquisition procedure. 
There is an additional benefit that there is no loss of light at the slices (`slit 
losses' in a conventional long slit spectrometer). The MRS slice width is 
set to be less than or equal to the FWHM of the diffraction limited point 
spread function at the slicing mirror. An equivalent long-slit spectrometer 
would vignette the light outside this region, losing about 50{\%} of the 
total. The amount of lost light could be reduced by making the slit wider, but
doing so would in turn reduce the spatial and spectral resolution and could also decrease the
signal to noise because of increased background radiation.

Second, from a scientific perspective, the wavelength range covered by the 
MIRI spectrometer is sufficiently broad (a factor of nearly six) that 
different emission mechanisms may dominate in different regions of the 
spectrum. In cases where these mechanisms do not share a common centre 
(e.g., stellar output compared with infrared re-emission in a starburst 
galaxy), a simple slit spectrograph poses a dilemma in placing the "source" 
on the slit. An IFU implements 3D spectroscopy, which solves this problem by giving accurately registered spatially
resolved spectroscopy over the entire field.

These considerations, combined with the wavelength coverage and resolving 
power requirements defined by the JWST mission science goals, and the mass, 
volume and electrical power limitations set by the JWST spacecraft 
environment, have resulted in a system with the characteristics summarised 
below and illustrated in block diagram form in 
Figure 1.

As shown in the figure, the full 5 to 28.5 $\mu $m wavelength range is 
divided within the Spectrometer Pre-Optics (SPO) into four simultaneous 
spectral channels \citep{kroes2005, kroes2010}, each with its own IFU \citep{lee2006} and using a simple scheme of 
pass-band separation by dichroic filters whose design is described in 
Hawkins et al. (2007) and Wells et al. (2004). We denote the short and long 
wavelength limits of each channel as $\lambda_{\mathrm{short}}$ and 
$\lambda_{\mathrm{long}}$. Each channel serves an optimised spectrometer. 
This separation provides several benefits: (1) it enables the use of diffraction 
gratings in first order, allowing each to be used near peak efficiency around the blaze wavelength, and (2) 
it also allows the IFU slice widths to be tailored to the wavelength-scaled FWHM in each channel. 

Electrical and thermal constraints limit the MRS to two 1024 x 1024 Si:As 
detectors (Ressler et al., 2014, hereafter Paper VIII), resulting in the 
spectrometer being split into two sets of optics, each with its own detector 
array, one for the two short-wave channels and one for the two long-wave 
channels. In each case, the spectra of two wavelength channels are imaged 
simultaneously onto the left and right halves of a detector array.

The product of spatial and spectral coverage that can be achieved in a 
single instantaneous exposure is ultimately set by the number of detector 
pixels. For the MRS, this results in a single exposure providing a spectral 
sub-band that covers only one third of each channel. Full wavelength 
coverage then requires three exposures, with a pair of mechanisms being used 
to select the gratings and dichroics in each case. We refer to the short and 
long wavelength limits of each sub-band as $\lambda_{\mathrm{min}}$ and 
$\lambda_{\mathrm{max}}$.
This design choice allows the grating performance to be optimised over a 
narrow wavelength range ($\lambda_{\mathrm{max}}$/$\lambda 
_{\mathrm{min}}$ $\sim$1.2). The gratings only have to work over 
20{\%} of the 1$^{\mathrm{st}}$ order near the blaze wavelength and the 
dichroics do not have to be used near the cross over between reflection and 
transmission. 

We will now step through the optical train as shown in 
Figure 1: the Input Optics and Calibration (IOC) 
modules pick off the MRS FOV from the JWST focal surface and pass it on to 
the SPO. The SPO spectrally splits the light into the 4 spectrometer 
channels and spatially reformats the rectangular fields of view into slits 
at the entrance of the Spectrometer Main Optics (SMO). The SMO comprises 
fixed optics that collimate the light for presentation to diffraction 
gratings mounted in the SPO and then re-images the dispersed spectra onto 
the two focal plane arrays. Key features of this optical system are 
described individually in the following subsections. The spectral and 
spatial coverage of the MRS is summarised in Table 
1.

\subsection{Telescope Focal Plane to Integral Field Units}
\label{subsec:mylabel1}
The field of view of the MRS is adjacent to the MIRI imager field and picked 
off from the JWST focal surface using the MIRI Pickoff Mirror (POM), which is 
common to both (see Paper II). The sky and the JWST pupil are reimaged by the IOC so that 
there is a pupil image at the SPO input and a sky image further on. A cold 
stop that is 5{\%} oversized with respect to the JWST pupil is placed at the 
SPO input pupil for straylight control.

The light is directed towards the first dichroic filter via a fold mirror 
placed 10 mm beyond the pupil. The focal plane is formed 535 mm beyond the 
pupil, providing the long path length and narrow beam waist needed between 
the pupil and the inputs of the four Integral Field Units (IFUs), for 
mounting a chain of dichroics to divide the light among the four 
spectral channels. 

A hole in the fold mirror, sized to be smaller than the footprint of the 
telescope central obscuration, acts as an aperture for the injection of 
light from the on-board Spectrometer Calibration Unit (SCU), shown in 
Figure 2. 
The SCU provides spatially and spectrally uniform blackbody illumination for 
flux calibration and pixel flat fielding functions, using as its light 
source a tungsten filament heated to a temperature of $\sim$1000 K 
by the application of an 8 mA drive current. The filament is mounted 
inside a non-imaging flux concentrator that generates spatially uniform 
focal plane illumination at the exit port of a 25 mm diameter reflective 
hemisphere, (described in Glasse et al., 2006). 
Two cadmium telluride lenses within the SCU then re-image the exit port onto 
the IFU input focal planes. The lenses provide a pupil image that coincides 
with the hole in the input fold mirror. This hole is positioned to lie 
within the footprint of the central obscuration of the JWST primary mirror 
and so has no impact on the science beam. In this way, the SCU can provide 
flood illumination of the full MRS field of view without any need for 
mechanisms or additional optical elements. 

The overall layout of the dichroic assembly is shown in 
Figure 3, with the input fold mirror and SCU 
situated at the left-hand end (but not shown). The three dichroics needed to 
divide the spectral band among the four spectrometer channels for one of 
the three sub-bands are indicated as D1, D2, and D3 in 
Figures 1 and 3.
Taking Sub-band A as an example, the required reflective band for dichroic 
D1 is the wavelength range for Sub-band A in Channel 1, while its 
transmission band needs to extend from the short end of Sub-band A in 
Channel 2 to the long end of Sub-band A in Channel 4. The bands 
are listed in Table 2 for all nine dichroics. In all cases, the 
mean reflectivity is above 0.95 and the mean transmission is above 
0.74.

All of the gratings work in first diffractive order so additional 
blocking is needed to reject second and higher orders. Because the dichroics 
work in series it is possible to use the combined blocking of dichroics 1 
and 2 to remove the need for blocking filters in Channels 3 and 4. For 
Channels 1 and 2, dedicated blocking is provided by the fixed filters 
shown as BF1 and BF2 in Figure 3, with light traps 
LT1 and LT2 absorbing unwanted reflections.

The path length required to reach the input of Channel 4 is greater than the 
535 mm discussed above, so the light transmitted by the final dichroic is 
re-imaged via an intermediate focal plane to the entrance of the Channel 4 
IFU.

The 21 mm diameter dichroic filters are mounted on two wheels. First is the 
nine-sided wheel A containing the three Channel 1 dichroic filters and three 
flat mirrors to direct the light towards Channel 1. The second, six-sided 
wheel, contains the six dichroics needed to divert the light into Channels 2 
and 3.
Each dichroic filter is mounted onto a diamond machined facet on the wheel 
that provides the required alignment accuracy and reduces the magnitude of 
any print-through of surface form errors from the wheel to the filter 
substrate. The filters are held in place with a spring loaded bezel with a 
clear aperture of 17 mm compared with the coated area of 17.4 mm. The bezel 
prevents light from reaching the uncoated area of the filter, which might 
result in un-filtered light entering the optical path. The blocking filters 
are mounted directly to the SPO chassis with a stand and bezel, similar to 
the mounting arrangement for the wheels.
The mechanisms that carry the dichroic wheels also carry 
the corresponding wheels with diffraction gratings for each 
sub-band (three per channel), as discussed in Section 2.4. By arranging for 
the gratings to be mounted on the same mechanisms as their band-selecting 
dichroics, the number of moving mechanisms within the MRS is kept to two. 

\subsection{The Integral Field Units}
\label{subsec:mylabel2}
Before the light in the four channels reaches the gratings it is 
sliced and reformatted by the Integral Field Units and so we describe these 
next. The parameters pertaining to the spatial and spectral coverage and sampling 
of the spectrometer, which are largely determined by the design of the four 
IFUs, are given in Table 1. Spatially, the image is 
sampled in the dispersion direction by the IFU slicing mirrors and in the 
slice direction by the detector pixels. Spectrally, the width of the slices 
defines the spectrometer entrance slit and the width of the image (in 
pixels) of the slice at the detector defines the width of the spectral 
sample. Inspection of Table 1 then shows that 
Channels 1 to 3 are all slightly undersampled spectrally, with slice widths 
less than 2 pixels at the detector. 

The bandwidths of the spectral channels listed in 
Table 1 were chosen such that the ratio $\lambda 
_{\mathrm{long}}$/$\lambda_{\mathrm{short}}$ in all four channels was 
the same: $\lambda_{\mathrm{long}}$/$\lambda_{\mathrm{short}} =$ 
(28.3 / 5)$^{\mathrm{1/4}} =$ 1.54. This means that, if each channel is assigned the same number of spectral pixels, the resolving power in each will be approximately the same. Each sub-band then has a width that is 
slightly larger than one third of the band-width of its parent channel, where the 
excess provides sufficient overlap of adjacent spectra to allow their 
concatenation. In practice, the overlap is typically 10 to 15{\%} of the 
spectral range, depending on the specific sub-bands and position in the 
field of view. 

For full Nyquist sampling of the telescope PSF the four IFU slice widths 
should ideally be matched to the half width at half maximum intensity of the 
PSF at the shortest wavelength in each channel. For the JWST telescope pupil 
this would equate to a$_{\mathrm{diff}} =$ 0.088 ($\lambda 
_{\mathrm{short}}$ / 5 $\mu $m) arcsec. However, the large increase in the 
beam size produced by diffraction at the IFU slicer mirror would then necessitate 
large apertures for the spectrometer optics to avoid vignetting. 
To control the size, mass and optical aberrations of the spectrometer, the 
starting point for determining the slice widths was therefore to set them equal to 2a$_{\mathrm{diff}}$, with fully sampled PSFs to be achieved using two or more 
pointings of the telescope.

Optimum spatial sampling, which minimises the number of telescope pointings 
needed to fully sample all four channels simultaneously in the across-slice 
direction, is  achieved by having a single pointing offset equal to 
n$_i$+1/2 slice widths for all channels where n$_i$ is a different integer for each channel. This led us to adopt a set of slice widths that follow a scheme 
where the slices are factors of 1, 11/7, 11/5 and 11/3 wider than the 
narrowest 0.176 arcsec slice. These values closely match the increasing FWHM of the PSF in the 4 channels and allow full Nyquist sampling to be 
achieved using a single `diagonal' offset whose magnitude in the across 
slice direction is equal to 11/2 times the width of the narrowest slice 
(0.97 arcseconds) and (N + 1/2) detector pixels in the along 
slice direction (where N is an integer).
With these slice widths fixed the design was interated to, as far as possible, meet the following criteria:  1.) each channel would occupy half a 1024 x 1024 pixel detector with borders to allow for tolerances in alignment and gaps between slices to avoid crosstalk; 2.) the FOV of each channel should be as large as possible and approximately square; and 3.) the long and short wavelength arms of the spectrometer should be identical (in practice one is the mirror image of the other). This iteration resulted in the parameters in Table 1.

The IFU design was developed from the IFU deployed in UIST, a 1-5 $\mu $m 
spectrometer for UKIRT \citep{ramsay2006}. Other examples of
all-reflecting image slicers are the SPectrometer for Infrared FaintField Imaging 
(SPIFFI) \citep{tecza2000}, and the slicer in the Gemini Near-Infrared Spectrograph 
(GNIRS) \citep{dubbeldam2000}.
The MIRI (and UIST) IFU design allows excellent control over stray light by providing
through-apertures for baffling. It consists of an entrance 
pupil, an input fold mirror, an image slicer mirror, a mask carrying exit pupils for the individual
sliced images, a mask carrying slitlets for the individual images, and an array of re-imaging mirrors
behind the slitlets. These components (except for the last) are shown in
Figure 4. The design is all-reflecting and constructed entirely of aluminium, and hence
is well-suited for operation in the infrared and at cryogenic temperatures, as discussed in Paper II. 
\citet{todd2003, lee2006}
describe the manufacture of the slicer. 

The optical path through the IFUs begins with the four toroidal mirrors, 
which comprise the anamorphic pre-optics (APO) module, (not shown in 
Figure 4). The APO re-images an area of up to 8 by 8 
arcseconds of the input focal plane onto the image slicer mirror, with 
anamorphic magnification. The image slicer (at the relayed image
plane) consists of a stack of thin mirrors angled to divide the image along the dispersion direction of the spectrometer 
into separate beams. In the across slice (dispersion) direction one 
slice width is matched to the FWHM of the Airy pattern at the shortest 
operating wavelength for the IFU.  In the along slice direction the 
magnification is chosen to provide the required plate scale at the detector. 
These two magnifications are not the same and so the APO exit pupil is 
elliptical, as illustrated in Figure 5 for Channel 3, where the 
footprint of the JWST pupil is shown in blue. 
The light exits the IFU through
individual pupil masks for each beam, then through individual slitlets (see Figure 4). Re-imaging mirrors
 behind the slitlets relay the beam to the input of the appropriate spectrometer. 
That is, each IFU takes the rectangular FOV and transforms it into a pattern of slitlets 
(as shown in Figure 4) that are re-imaged onto the entrance aperture 
for its corresponding grating spectrometer. 

The slicing mirror comprises a number of spherical optical surfaces, diamond 
turned onto a common substrate. This manufacturing approach was used previously for
the IFU in GNIRS  \citep{dubbeldam2000}. Figure 6 is a 
photograph of the image slicer for Channel 1, which has 21 slicer 
mirrors arranged about the centre of the component with 11 slices on one 
side and 10 on the other. Each mirror is offset to direct its output beam 
towards the corresponding re-imaging mirror. The design for channels 2 to 4 
is similar but with correspondingly fewer and larger slices as the wavelength 
increases, as listed in Table 1. The mirrors are 
arranged in a staircase-like manner to aid manufacturing.
Each of the slices re-images the IFU entrance pupil at a scale of $\approx 
$1:1 and because the centres of curvature of the mirrors are offset 
laterally with respect to each other, the pupils that the slices produce are 
separated, as can be seen in Figure 4. 

The slicing mirrors truncate the anamorphically magnified image of the sky 
along the dispersion direction, as shown in Figure 
7 (a) to (c). This truncation results in the pupil image being diffraction 
broadened in the dispersion direction as shown in 
Figure 7 (d). To reduce vignetting, the baffle at 
this pupil image and subsequent optical components (including the 
diffraction gratings) are oversized. This is illustrated in 
Figure 7(d) where the geometric footprints for 3 
field positions are overlaid along a slice to appear as the blurred green 
patch in the image. The notional, geometric pupil is shown in red and the 
oversized mask aperture is shown as a green rectangle. 

The re-imaging mirror array forms 
images of the slicing mirrors as input slits of the spectrometer and also 
images the pupils at infinity, making the output telecentric. Finally, a roof 
mirror redirects light from the re-imaging mirrors so that the output 
consists of the two lines of staggered slitlets relayed from those seen in 
Figure 4.
A slotted mask at the position of the IFU output slits defines the field of 
view along the individual slices, while in the spectral direction the slots 
are oversized to ensure they do not vignette the image, but still act as  
baffles to reduce scattered light between the slices. 

The numbers of detector columns covered by the four IFUs are approximately 
476, 464, 485, 434 for channels 1 to 4 respectively. The size of the gap 
between individual slits for channel 1 is set to be approximately equal to 
the diameter of the first dark diffraction ring in the telescope PSF, (about 
4 pixels). For channels 2 -- 4 the size of the gap is slightly greater than 
the diameter of the first dark ring.

\subsection{The Spectrometer Main Optics (SMO)}
\label{subsec:mylabel3}
The SMO (whose layout is shown in Figure 8), comprises four grating spectrometers
 in two arms. The development and test of these spectrometers are described 
by \citet{kroes2005, kroes2010}.  Each of them performs three functions: collimation of the telecentric 
output beams of one of the four IFUs, dispersion of the collimated beam, and imaging of 
the resulting spectrum onto one half of one of the two focal plane arrays. 
One of the two spectrometer arms includes the two short wavelength channels 
(1 and 2), and the other the long wavelength channels (3 and 4). 

We will now 
step through the optical path from the IFUs to the 
detectors. The  IFU output beam for each channel is collimated and its light directed towards the corresponding diffraction gratings as shown in Figure 9. Each spectrometer arm uses 6 gratings (two wavelength channels, three sub-band exposures). Figure 8 then shows the optical paths from the gratings to the detectors. The dispersed beams are imaged by three-mirror-anastigmat (TMA) camera systems (M1-M2-M3). In each spectrometer arm the channels have separate, but identical, M1 camera mirrors that provide intermediate images of the spectra between M1 and M2. Folding flats at these intermediate focus positions reflect the Channel 1 and Channel 4 beams such that the combined (Ch. 1+2) and (Ch. 3+4) beam pairs are imaged onto opposite halves of the detectors by common M2 and M3 mirrors. The symmetry through the centre plane of the camera optics and the positioning of the optics allow for an opto-mechanical design of two mirror-imaged boxes, with identical optics and almost identical structures. The combination of symmetry and an all aluminium design (optics as well as structures) allowed the very strict alignment and stability requirements to be met fully with 
an adjustment-free mounting (i.e. positional accuracy by manufacturing accuracy only  except for the focus shim at the detector interface), a major advantage regarding the total design, manufacture and test effort.

The mirror substrates are all light-weighted aluminium with diamond turned 
optical surfaces that are gold coated for maximum reflectivity at the MIRI 
operating wavelengths. They are mounted through holes in the housing walls 
using stress-relieving lugs under light-tight covers to prevent any high 
temperature radiation from the instrument enclosure (at $\sim$40 
K) reaching the detectors. The two sets of gratings for the A, B and C sub-bands of the two channels are mounted on a single wheel, as shown in figure 10, that is mounted on a mechanism  that also carries the dichroics wheel on a lower level. The mechanisms are located in the SPO and each mechanism is rotated by a single actuator to give the correct combination of dichroics and gratings across all channels for each exposure. The gratings are master rulings on aluminium substrates and are gold coated. Table 3 lists the design parameters of the gratings. All gratings operate in first order. The angular values quoted in Figure 1 are calculated for the (virtual) nominal input beam, originating from the centre of the IFU output area.

\subsection{Stray Light Control}

The SPO is designed to be a well baffled optical system. 
To intercept stray light from the telescope the first optical element in the 
MRS is a cold stop placed at the entrance pupil.  A number of 
features within the MRS help to control stray light: 1.) stops at each pupil and sky image in 
the optical train; 2.) light traps where significant stray light occurs 
(0$^{\mathrm{th}}$ order from the gratings); 3.) labyrinths at the edges of 
apertures between modules and tight control of all external apertures, e.g. 
electrical feed throughs; 4.) black coating of all surfaces not in the main optical path, which is an inorganic black anodising applied to a roughened surface of aluminium. The process was carried out by Protection des Metaux, Paris; and  5.) avoidance of surfaces at grazing incidence near the 
main optical path. In addition, for each slice of each IFU there are output 
pupil and slit masks.

All the diamond turned aluminium mirrors in MIRI were specified to have a 
surface roughness of \textless 10nm RMS. This results in a total integrated 
scatter per surface of \textless 0.06{\%}; non-sequential optical modelling 
indicated that this level of scatter should not significantly degrade the 
PSF of the MRS. All mirrors have an 
overcoated gold surface with reflectivity \textgreater 98.5{\%} in the MIRI 
wavelength range. 

Straylight analysis shows that the extensive baffling combined with the low 
scattering optical surfaces and blackened structure should reduce unwanted light 
(cross talk between channels, degraded PSF and out-of-beam light at the 
output image) to levels that are estimated to be less than 0.1{\%} of the 
 background radiation included in the science beam.

\section{Measured performance}
\label{sec:measured}
\subsection{The Spectral Image}
\label{subsec:mylabel4}

Due to the nature of an IFU spectrograph with its slicing and dispersing optics, the resulting detector images are not straightforward to analyze: Spatial, spectral, and with it photometric information, are spread over the entire detector array. Further, the gratings are used in an optically fast, non-Littrow conﬁguration. The resulting anamorphic magnification varies from the short to long wavelength limit for each sub-band which, when combined with the curved spectral images of the dispersed slices on the detector as shown in Figure 11, makes the optical distortion significant and complex.
The anamorphic and slicing optics add other components of distortion that vary the plate scale in the along-slice direction for each slice individually. This leads to a very complicated variation of the spatial plate scales and finally to a dependency of the spatial and spectral axes on each other and on the location on the detector array.

To enable scientific studies based on MRS data, the flux measured in each 
detector pixel needs to be associated with a wavelength and location on the 
sky. Since the IFU provides two spatial dimensions, each detector pixel 
corresponds to a position within a three-dimensional cube with two spatial and one spectral 
dimension. Due to the optical distortion, the edges of these cubes are 
neither orthogonal nor constant in length. Consequently, we have developed a 
process (called image or cube reconstruction), which allows a transformation 
of the detector pixels onto orthogonal cubes.

Cube reconstruction requires a thorough understanding of the optical 
distortion. It is not possible to just characterize the curved images of the 
dispersed slices (as shown in the left plot of 
Figure 11) by approximating the curvature with 
polynomial functions. Due to the distortion caused by the slicing optics, 
the spatial plate scale varies within a slice non-uniformly. The 
characterization of the plate scale and the association of each detector 
pixel with its projected location on the sky at any given wavelength is a 
task for the astrometric and wavelength calibration, described in the next 
sections. Once the sky-coordinates and wavelength are known for each 
detector pixel, the cube can be reconstructed as discussed by Glauser et. 
al. (2010).

\subsection{Astrometric Calibration}
To describe the optical distortion on the image plane of the detectors we 
began from the optical model, using ray-tracing techniques to project each 
detector pixel through the slicing optics backwards onto the sky. When the 
as-built geometry was taken into account, we achieved a high spatial 
accuracy, as demonstrated for the MIRI Verification Model (Glauser, et al., 
2010). To verify this approach, a dedicated astrometric calibration campaign 
was conducted during the instrument test campaign of the MIRI Flight Model. 

As outlined in Glauser et al. (2010), the intra-slice spatial distortion can 
be approximated with a 2$^{\mathrm{nd}}$ order polynomial to accuracies of a 
few milli-arcseconds - much better than required for the astrometric 
accuracy given the minimum plate scale of 0.196 arcseconds in the 
along-slice direction for channel 1. We conducted the astrometric 
calibration by placing a broad-band point source at three field positions 
for each slice and each channel and recording the dispersed signature on the 
detector. The central pixel of the spatial profile was determined and 
correlated with the known position of the point source (our reference on the 
sky). With this method we were able to determine the spatial plate scale at 
any location on the detector. Due to limitations of the test setup, i.e., 
strong field distortion of the steerable point-source, the achievable 
relative astrometric accuracy was very limited. However, within an estimated 
upper limit of $\sim$0.5 pixels (0.098~arcsecond for channels 1 
and 2, 0.123~arcsecond for channel 3, and 0.137~arcsecond for channel 4), 
the approach of using the optical model for the reconstruction was 
validated.

A further result from this calibration campaign was the measurement of the 
FOV for each sub-band. Figure 12 shows the MRS FOV 
in the JWST coordinate frame and its position relative to the imager field. 
The FOV common to all MRS channels is 3.64~arcsec in the along-slice 
($\alpha )$ and 3.44~arcsec in the across-slice direction ($\beta )$.

\subsection{Wavelength Calibration}
To determine the absolute wavelengths for the MRS channels we conducted a 
series of calibration measurements during the instrument test campaign. 
Fabry-P\'{e}rot etalon filters were used to create a dense pattern of 
unresolved spectral lines on the detector. Many tens of lines were typically 
visible at a signal to noise ratio of more than 50 in a single sub-band 
exposure. An example etalon measurement is shown in Figure 13. To provide a reference wavelength to distinguish between adjacent 
etalon lines, the telescope simulator used in the test was also equipped 
with a pair of `edge' filters, one of which cut-on at around $\lambda = $ 
6.6 $\mu $m (in Sub-band 1C) and the other which cut off at $\lambda =$ 
21.5 $\mu $m (in Sub-band 4B). 

The absolute wavelengths of the etalon lines and edge filters had previously 
been measured at ambient and 77 K temperatures using a laboratory standard 
Fourier Transform spectrometer with a spectral resolving power of R $=$ 
100,000. The extrapolation of the wavelength scale to the 34 K operating 
temperature for the test campaign was based on published measurements 
(Browder and Ballard, 1969, Browder and Ballard, 1972 and Smith and White, 
1975). This extrapolation resulted in a typical wavelength correction of 
less than 1{\%} of the width of the MRS spectral resolution element. 
 The repeatability of the scale after multiple mechanism reconfigurations 
has been measured to be 0.02 resolution elements.  

The wavelength calibration process then involved the assignment of an 
absolute wavelength to an etalon line by fitting the calibrated spectra of 
the edge filters to their MRS measured spectra in Sub-band 1C and 4B. The 
known separation of the etalon lines was used to extend the wavelength scale 
across the full spectral image in each of these sub-bands. 
To extend the scale to other sub-bands, 
pairs of measured etalon spectra were co-added and the positions of unique 
identifying features (due to spectral beating between the two patterns) were 
used. A set (2 per sub-band) of second order polynomial fits to the 
positions of the etalon lines was used to generate a wavelength value for 
the corners of all illuminated pixels. 

The wavelength calibration derived in this way was encoded for use in the 
MRS calibration pipeline by forming six images (one for each MRS detector 
and all three grating wheel settings) where the image signal values were set 
equal to the wavelengths at the corner of each detector pixel. We note that 
as a result, these wavelength reference images have one more row and column 
than the detector images. 
The relative accuracy of this wavelength scale (within and between 
sub-bands) is estimated to be better than 0.02 spectral resolution elements 
which, for example, corresponds to 0.03 nm at $\lambda =$ 5 $\mu $m. The 
absolute accuracy of the wavelength scale is estimated to be comparable to 
this figure but this will need to be confirmed during on-orbit commissioning 
using spectral standards. 

The wavelength calibrated etalon spectra were also used to measure the 
spectral resolving power of the MRS. The results are shown in 
Figure 14, where the sub-band averaged values are 
shown in red and the spread of values seen across the field of view is 
indicated by the black band.
We note that Figure 14 does not take account of the 
intrinsic spectral width of the etalon features, which was determined using 
the R $=$ 100,000 calibrated spectra, described above. Initial efforts to 
deconvolve the intrinsic line profiles from the MRS measured spectra suggest 
that the resolving powers quoted in Figure 14 may 
be underestimated by around 10 {\%}. We have therefore used the band 
averaged measurements scaled by a factor 1.1 to generate the summary values 
quoted in Table 1.

\subsection{Spatial Sampling }
\label{subsec:spatial}
As outlined in Section \ref{sec:optical}, the PSF of the MRS is 
under-sampled by design, with full sampling in both spatial and spectral 
dimensions requiring that the object be observed in at least two dither 
positions that include an offset in the across-slice direction of 11/2 times 
the channel 1 slice width (which corresponds to 7/2 slices in channel 2, 5/2 
slices in channel 3, and 3/2 slices in channel 4). Due to the curved shape 
of the distorted spectrum on the detector and the variable plate scale along 
the individual slices, the exact dither offset in the along-slice direction 
is less well determined (but also less critical). Figure 15 shows the nominal MRS 
dither pattern to be used in a single observation to sample point sources 
fully, as derived during test campaigns, and as proposed for in-flight 
operations. 

To achieve a fully sampled PSF, these dithered observations must be 
combined. We anticipate that this could most readily be achieved using the 
reconstructed cubes. To avoid loss of spatial resolving power caused by any 
shift- and co-adding algorithm due to the re-binning of the data (for 
example, Fruchter {\&} Hook, 2002), we minimize the necessary re-binning 
steps by incorporating dither offsets parallel to the slice into the 
reconstruction algorithm itself. This has the advantage that only one 
re-binning step is required from detector data to reconstructed cubes, while 
dither offsets in the across slice direction can be corrected and combined 
afterwards using an interlacing method. We expect more 
sophisticated techniques than we have developed so far 
to be incorporated into the data reduction pipeline before 
launch.

\subsection{PSF and Optical Quality}
\label{subsec:mylabel5}
We attempted to measure the MRS PSF during flight model testing at RAL using 
the MIRI Telescope Simulator (MTS), described in Paper 
II. However, optical aberrations and vignetting in the MTS led to the 
generated point source being elongated and extended at short wavelengths, 
such that it did not provide a sufficiently point-like image. Even at longer 
wavelengths, where aberrations became less apparent as diffraction started 
to dominate, the measured PSF was broader than nominal due to vignetting in 
the telescope simulator. 

We therefore repeated the PSF measurement on the flight model during the 
first cryo-vacuum test campaign at NASA-Goddard (CV1RR), where a compact and 
well defined point source was available. Deep exposures at 17 different 
locations in the MRS field were combined and reconstructed to form the PSF 
image for Channel 1A, shown in Figure 16. For these 
data, the spectral coverage was limited to a narrow (0.125 $\mu $m wide) 
wavelength range around 5.6~$\mu $m. 
A comparison with the model PSF (pure diffraction limited 
Fourier-transformed JWST pupil) shows a very good match across the slices. 
In the along-slice direction, a broadening of approximately 50{\%} is observed. 
Currently, possible causes considered for the broadening include scattering 
in the detector substrate (the detector halo effect, which is also observed 
in the imager and is discussed in Rieke et al., (2014, Paper VII) or a 
side-effect of the straylight discussed in Section 
\ref{subsec:mylabel6}. In the case of the halo effect, the larger 
degree of broadening seen in the MRS may be due to the larger spatial sample 
per pixel. More detailed modelling is required to confirm the root cause.

\subsection{Spectral Fringing}
\label{subsec:spectral}
Spectral fringes are a common characteristic of infrared spectrometers. They 
originate from interference at plane-parallel surfaces in the light path of 
the instrument. These surfaces act as Fabry-P\'{e}rot etalons, each of which 
can absorb light from the source signal with a unique fringe pattern. In the 
infrared wavelength range, surfaces separated by a fraction of a mm up to a 
few cm may form very efficient etalons. The most obvious source of fringes 
in the MRS is the detector itself with a physical thickness of 500 $\mu $m 
(Paper VII). Similar fringes have also been observed in spectra measured 
with the MIRI Verification Model during testing and with the Spitzer-IRS 
instrument, which employs comparable (though smaller) Si:As BIB detectors, 
Lahuis {\&} Boogert (2003).

For the initial analysis we follow the formalism as defined in Kester et 
al., (2003) and Lahuis {\&} Boogert, (2003). The sine approximation, $I_{t} = A_{f}$\textit{ cos(2}$\pi 
\omega D) + B_{f}$\textit{ sin(2}$\pi \omega D)$, is used with $\omega $ being the 
wavenumber and $D$ the optical thickness of the instrument component. Of 
primary interest to help in the identification and first characterization of 
the fringes is the optical thickness $D$. In the MRS test data three distinct 
fringe components are seen, with key parameters listed in 
Table 4.

Of the three fringe components two are directly matched to optical components in the instrument.  
The main fringe component (\#1 in Table 4) has a derived optical thickness of approximately 3.5 mm 
for all sub-bands.  This corresponds to the optical thickness of the detector substrate which has d $\sim$ 500 $\mu$m 
and n $\sim$ 3.42, giving D $\sim$ 0.34 cm.  Figure 17 gives an illustration of this fringe pattern, 
showing the main detector and dichroic fringes in more detail for Channel 4.
The optical thickness derived for the second, high frequency, component (D $\sim$ 2.7 cm) is matched to that of the CdTe Dichroic filters (Section 2.2).  For the third set of low frequency components (D = 0.01 to 0.1 cm), no unique surfaces in the instrument are identified; instead these are likely to originate from beating between primary fringe components.  The fringe variations come from the layered structure of the detector substrate and optical thickness differences between individual dichroic filters.

Figure 18 shows the peak normalized fringes over the
entire wavelength range of the MRS. The two curves in 
Figure 18 show predictions for the expected fringe 
amplitudes based on representative anti-reflection (AR) coating profiles as 
applied to the MIRI flight detectors. The solid curve assumes a pure 
two-sided etalon while the dashed-dotted curve simulates a back-illuminated 
surface with a fully reflective front surface and photon absorption in the 
active layer (adopted from Woods et al., 2011). Though not a perfect match, 
this approximate detector model does reproduce the general trend and 
magnitude of the fringes.

Fringe removal will be achieved using the techniques developed for and 
applied to ISO and Spitzer data (see Lahuis {\&} van Dishoeck, 2000; Kester 
et al., 2003; Lahuis {\&} Boogert, 2003). This involves dividing the 
observed spectra by a fringe flat-field followed by the removal of fringe 
residuals using the sine fitting method. This approach has proven to be 
reliable and robust for most spectra and the experience with ISO and Spitzer 
has shown that it allows the removal of fringe residuals down to the noise 
level. The main limitations with this technique are the definition of the 
spectral continuum in the presence of spectral features and isolating the 
fringe spectrum from broad molecular (vibration-)rotation bands (e.g. those of C$_2$H$_2$, 
HCN, CO$_2$ and H$_2$O). 

For the MIRI IFU other effects play a role and may limit the fringe removal 
for point and compact source measurements. The two major effects are; i) the 
illumination (and its effective angle) on the detector depends on the source 
morphology (position and extent) and ii) the wavelength depends on the 
spatial location of the point source in the IFU field. This results in i) 
changes in the effective optical thickness from source to source and ii) a 
wavelength shift with spatial offset. Both modify the detailed fringe 
pattern for individual cases. Figure 19 illustrates 
this with point source measurements from the CV1RR test campaign and using 
extended source blackbody spectra from flight model testing at RAL. Small 
sub-pixel pointing offsets are seen to have a discernable impact on the 
fringe pattern.

These effects can be mitigated by traditional fringe removal techniques 
using optimized and iterative reduction algorithms (e.g. by modifying, 
shifting and stretching the reference fringe spectrum before applying it). 
This has been used in individual ISO and Spitzer cases and the MIRI team 
will work on developing optimized methods for the MRS. 
For MIRI an alternative model-based approach is also under study which uses the 
observed source morphology to define the fringe spectrum. This method will 
be applicable to both the MRS and the LRS, and will complement the 
traditional fringe removal techniques. This requires both a well-defined and 
well-calibrated fringe model (an ongoing MIRI team activity) and a flexible 
and iterative reduction pipeline (in development at STScI based on input 
from the MIRI team).

\subsection{Pixel flat field}
\label{subsec:pixel}
We determined the pixel-to-pixel variation of the response for both short 
and long wavelength detectors, using measurements taken during testing at 
RAL. The illumination was provided by an external, extended source in the 
MTS. A number of exposures measured over the period of the whole test run 
and covering the full MRS wavelength range were included in this 
calculation.

We first de-fringed the data using the prescription described in Section 
\ref{subsec:spectral}, with the results illustrated in Figure 17. We then calculated the mean value of 
each pixel, $f_{ave}(i,j)$, by averaging over 5 x 5 pixel boxes centered on pixel 
($i,j)$ and accounting for pixels close to the edges of the slices or near bad 
pixel clusters. The original data were then divided by these averaged values 
to create a map of the normalized pixel gain. We checked the distribution of 
the pixel-to-pixel variation for each exposure in all the available data for 
both detectors and all the sub-bands and found that the distribution is 
Gaussian with a full-width-half-maximum of $\sim$2{\%} 
(Figure 20). The variation among different 
datasets has a standard deviation of $\le $ 0.2{\%} for 16 different 
observations. The pixel-to-pixel variation does not appear to be wavelength 
dependent as both short and long wavelength detectors show the same overall 
flatness. This uniformity suggests that the correction of pixel response 
variations on orbit can be achieved using infrequent calibration 
measurements.

\subsection{Spectrophotometric Performance}
\label{subsec:spectrophotometric}

The absolute responsivity of the MRS is expressed in terms of the quantity 
referred to as the Photon Conversion Efficiency (PCE), which is equal to the 
number of electrons detected by the focal plane array for each photon 
incident at the MIRI entrance focal plane. 
The wavelength-dependent PCEs for the MRS were derived during testing at 
RAL. This was done by configuring the MTS to provide extended illumination 
of the MIRI entrance pupil with the spectral energy distributions of 
blackbodies of 400, 600 and 800 Kelvin. For each MTS blackbody configuration, 
MRS spectra were obtained in all 12 spectral bands, together with background 
measurements using the blank position in the MTS filter wheel.

The data were processed using the standard DHAS tool (Paper II), which 
converts the raw readouts of the integration ramps to slopes in physical 
units (electrons/sec). The DHAS miri\textunderscore cube module, which performs the 
reconstruction technique described in Section \ref{subsec:mylabel4}, 
was then used to construct spectral cubes from the slope images, re-gridding 
the focal plane array pixel signals onto an equidistant spectral cube. The 
spectral cubes of the background measurements (MTS filter blank position) 
were subtracted from the flood illumination measurements to correct for the 
test facility background.

The spectral cubes produced with the DHAS miri\textunderscore cube routine 
have a fully calibrated WCS with the plate scale and wavelength coverage of 
every cube pixel. When combined with an estimate for the absolute flux 
delivered to the entrance focal plane by the MTS (Paper II), this allows us 
to calculate the photon conversion efficiency in every pixel of the spectral 
cube for all MRS spectral bands. Table 5 lists the 
mean PCEs for all MRS bands. We estimate the fractional error to be 20 {\%} 
in these figures, due to systematic effects, primarily in estimating the 
absolute flux from the MTS.

One obvious feature of Table 5 is the sharp drop in 
PCE from Channel 3 to Channel 4. The PCE in Channel 4 is roughly a factor of 2.5 lower than 
was expected from sub-system measurements of the nominal MRS optical train. 
The extra loss was identified (M. te Plate, ESA, private communication) as 
being caused by a fault in the groove profiles of the Channel 4 gratings, 
which cannot be corrected before launch. 

Following the procedure to determine the photon conversion efficiency, we 
can establish a first spectrophotometric calibration. By comparing the flux 
conversion factors derived from blackbody measurements of the MTS at 400, 
600 and 800K, we can assess the achievable spectrophotometric calibration 
accuracy. Figure 21 shows the ratios of the 
different obtained flux conversion factors as a function of wavelength. We 
are encouraged by the good agreement (less than 2 {\%} variation) over a 
large swathe of the MRS wave-band. 

\subsection{MRS Straylight}
\label{subsec:mylabel6}
As described in Sections 2.3 and 2.5, great care was taken to 
minimise sources of straylight and optical cross-talk within the MRS IFUs. 
However, a source of straylight was detected during RAL testing, which was 
identified as being caused by scattering in optical components within the 
SMO. The stray light is manifested as a signal that extends in the detector 
row direction. Its magnitude is proportional to that of bright illuminated 
regions of the spectral image, at a ratio that falls with increasing 
wavelength, from about 2 {\%} in Channel 1A to undetectably low levels 
longward of Channel 2B. 
\figurename~\ref{fig23} emphasises the impact of the straylight 
in Channel 1B, using the wavelength-averaged, reconstructed image of a 
bright source (seen at the top-left). The straylight signature is seen as 
the two horizontal bands, where the variation in brightness with the `alpha' 
coordinate is well explained by the mapping between alpha and detector row 
coordinate. 

The development of algorithms for correction of this straylight is underway. 
They take advantage of the stray-light being the dominant signal in the 
inter-slice regions of the detector, thereby making it amenable to accurate 
characterisation. Initial indications (Figure 22)  
suggest that an effective correction algorithm will be available before 
launch. 

\subsection{Spectral Leakage}
Testing at RAL revealed a gap in the performance of the train of dichroics 
described in Section 2.2. The set of filters that 
define the pass-band for Channel 3, Sub-band A, have an unwanted (but small) 
transmission peak at a wavelength of 6.1 $\mu $m, which allows light in the 
second diffraction order to reach the detector at the position where 12.2 
$\mu $m light is detected in the first diffraction order. This leak was 
confirmed using Fabry-P\'{e}rot etalon data and characterised to produce the 
leak profile plotted in Figure 23. This curve can 
be interpreted as the transmission profile by which the 6.1 
$\mu $m spectrum of a target object should be multiplied to determine the leakage 
signal at 12.2 $\mu $m. 

There are two options for mitigating the effects of the spectral leak. 
First, the Channel 3A spectrum can be corrected by multiplying an 
observation of the 6.1 $\mu $m (Channel 1B) spectrum of the target object by the 
leak profile, resampling the wavelength scale of the resulting spectrum to 
the wavelength grid of Channel 3A and then subtracting it from the 
contaminated Channel 3A spectrum. This requires that a separate Channel 1B 
observation has been taken. 

The second method would be to make the Channel 3A observation with Dichroic Wheel 1 set to use the Sub-band C  dichroics (see Section 2.2 to see what this means in terms of the optical train). This combination of dichroics reduces the spectral leak by a factor of more than 1000 at $\lambda$ = 6.1 $\mu$m. The unwanted side-effect of this solution is up to a  factor of three loss of PCE at the short wavelength ends of Channel 2A and Channel 4A.

\section{Conclusion}
We have presented the key parameters that describe the performance of the 
MIRI MRS spectrometer as designed or measured, in a form that both provides 
our best estimate of the behaviour of the instrument on-orbit and also that is 
accessible to the prospective user. The optical design behind the parameters 
is presented at a level of detail that is intended to provide the astronomer 
with an understanding of what to expect in terms of operating restrictions 
and data format when planning observations. 

The impact of straylight and spectral leaks in contaminating the spectral 
images has been discussed, along with proposals of operational and 
analytical techniques that should mitigate their effects. When combined with 
the latest sensitivity estimates (Glasse et al., 2014, Paper IX), we are 
confident that the MIRI MRS will meet all of its scientific objectives as 
part of the JWST Observatory.

\section{Acknowledgments}
The work presented is the effort of the entire MIRI team and the enthusiasm within the MIRI partnership is a significant factor in its success. MIRI draws on the scientific and technical expertise many organizations, as summarized in Papers I and II. 
A portion of this work was carried out at the Jet Propulsion Laboratory, California Institute of Technology, under a contract with the National Aeronautics and Space Administration.

We would like to thank the following National and International
Funding Agencies for their support of the MIRI development: NASA; ESA;
Belgian Science Policy Office; Centre Nationale D'Etudes Spatiales;
Danish National Space Centre; Deutsches Zentrum fur Luft-und Raumfahrt
(DLR); Enterprise Ireland; Ministerio De Economi{\'a} y Competividad;
Netherlands Research School for Astronomy (NOVA); Netherlands
Organisation for Scientific Research (NWO); Science and Technology Facilities
Council; Swiss Space Office; Swedish National Space Board; UK Space
Agency.

\eject

\begin{deluxetable}{lccccc}
\tabletypesize{\footnotesize}
\tablecolumns{6}
\tablewidth{0pt}
\tablecaption{Spatial and spectral parameters for the 4 MRS channels}
\tablehead{\colhead{    }             &
          \colhead{Channel}                       &
	\colhead{1}       &         
          \colhead{2}       &       
          \colhead{3}       &       
          \colhead{4}   \\
 	 }
\startdata
Slice width  & arcsec    &  0.176   &  0.277    &  0.387   &   0.645   \\
Number of slices    &  |  &   21  &   17  &   16  &  12  \\
~~~~~~~~~~~~~~~~~~~~~~~~~~~~~~ $\lambda_{short}^a$ & pixels  &1.405   &  1.452    &  1.629   &   2.253   \\
Slice width at detector &        &   &  &  &  \\
~~~~~~~~~~~~~~~~~~~~~~~~~~~~~~~$\lambda_{long}^b$  &  pixels  &  1.791   &  1.821   &  2.043   & 2.824   \\
Pixel size along slice  & arcsec/pixel   &  0.196    &  0.196   &  0.245   &  0.273    \\
FOV (across $\times$ along slices)  &   arcsec  &  3.70 $\times$ 3.70  & 4.71 $\times$ 4.52  &  6.19 $\times$ 6.14  &  7.74 $\times$ 7.95  \\
\hline 
     &      &  Sub-band A  &     &     &    \\
%\multicolumn{2}{lc|c|c|c|}{}
Wavelength range~~~~~~ $\lambda_{min}^c$ - $\lambda_{max}^d$   & $\mu$m  &  4.87 - 5.82   &  7.45 - 8.90 &  11.47 - 13.67  &  17.54 - 21.10  \\
Resolution  & $\lambda/\Delta \lambda$    &   3320 - 3710  &  2990 - 3110  &  2530 - 2880  & 1460 - 1930 \\
\hline
     &      &  Sub-band B  &     &     &    \\
Wavelength range ~~~~~~ $\lambda_{min}$ - $\lambda_{max}$   & $\mu$m  &  5.62 - 6.73   &  8.61 - 10.28  &  13.25 - 15.80  &  20.44 - 24.72  \\
Resolution  & $\lambda/\Delta \lambda$    &  3190 - 3750  &  2750 - 3170  &  1790 - 2640  & 1680 - 1770 \\
\hline
     &      &  Sub-band C  &     &     &    \\
Wavelength range ~~~~~~ $\lambda_{min}$ - $\lambda_{max}$   & $\mu$m  & 6.49 - 7.76   &  9.91 - 11.87 &  15.30 - 18.24  &  23.84 - 28.82  \\
Resolution  & $\lambda/\Delta \lambda$    &  3100 - 3610  &  2860 - 3300  &  1980 - 2790  & 1630 - 1330 \\
\enddata
\tablenotetext{a}{$\lambda_{short}$ is the shortest wavelength of operation for the corresponding channel.}
\tablenotetext{b}{$\lambda_{long}$ is the longest wavelength of operation for the corresponding channel.}
\tablenotetext{c}{$\lambda_{min}$ is the shortest wavelength of operation for the corresponding sub-band within a channel.}
\tablenotetext{d}{$\lambda_{max}$ is the longest wavelength of operation for the corresponding sub-band within a channel.}
\end{deluxetable}

\begin{deluxetable}{lcc}
\tabletypesize{\footnotesize}
\tablecolumns{3}
\tablewidth{0pt}
\tablecaption{Reflective and transmissive bands of the MIRI dichroic beam-splitters}
\tablehead{\colhead{Identification}             &
	\colhead{Reflection band ($\mu$m)}       &       
          \colhead{Transmission band ($\mu$m)}   \\
 	\colhead{}                                             &
          \colhead{R $\ge$ 94\% }                     &
          \colhead{T $\ge$  74\%}                     \\
} 
\startdata
Dichroic 1A  &  4.84 - 5.83     &  7.40 - 21.22 \\
Dichroic 1B  &  5.59 - 6.73     &  8.55 - 24.73 \\
Dichroic 1C  &  6.45 - 7.77     &  9.87 - 28.5 \\
Dichroic 2A  &  7.40 - 8.91     &  11.39 - 21.22 \\
Dichroic 2B  &  8.55 - 10.29   &  13.16 - 24.73 \\
Dichroic 2C  &  9.87 - 11.88   &  15.20 - 28.5 \\
Dichroic 3A  &  11.39 - 13.68 & 17.45 - 21.22 \\
Dichroic 3B  &  13.16 - 15.80  & 20.34 - 24.73 \\
Dichroic 3C  &  15.20 - 18.25  & 23,72 - 28.5 \\

\enddata
\end{deluxetable}

\begin{deluxetable}{cccccc}
\tabletypesize{\footnotesize}
\tablecolumns{6}
\tablewidth{0pt}
\tablecaption{MRS grating dimensions: groove length; ruled width; angle of incidence; and angle of diffraction. All gratings have a blaze angle of 45 degrees.}
\tablehead{\colhead{Band}             &
	\colhead{Grating constant}       &       
           \colhead{Length}      &       
           \colhead{Width}      &       
           \colhead{$\theta_{inc}$ }       &       
           \colhead{$\theta_{diff}$ }       \\       
           \colhead{}   &
	\colhead{(mm$^{-1}$)}       &       
           \colhead{(mm)}      &       
           \colhead{(mm)}      &       
           \colhead{(deg)}       &       
           \colhead{(deg)}       \\       
   	 }
\startdata
  &  & Channel 1   &  &  &  \\
 A  &  266.67  &  &  &  &\\
 B  &  230.77  &  28  &  44  &  55.46  &  29.2  \\
 C  &  200.00  &  &  &  &   \\
\hline
  &  & Channel 2   &  &  &  \\
 A  &  171.43  &  &  &  &\\
 B  &  148.45 &  30  &  63  &  54.46  &  28.5  \\
 C  &  128.57  &  &  &  &   \\
\hline
  &  & Channel 3   &  &  &  \\
 A  &  112.06  &  &  &  &\\
 B  &  96.97  &  30  &  63  &  54.46  &  28.5  \\
 C  &  83.96  &  &  &  &   \\
\hline
  &  & Channel 4   &  &  &  \\
 A  &  71.43  &  &  &  &\\
 B  &  61.28  &  34  &  64  &  53.46  &  27.2  \\
 C  &  52.55  &  &  &  &   \\

\enddata
\end{deluxetable}

\begin{deluxetable}{ccccc}
\tabletypesize{\footnotesize}
\tablecolumns{5}
\tablewidth{0pt}
\tablecaption{Observed fringe components with basic parameters and identification. The dichroic is modelled as chemical vapour deposited CdTe.}
\tablehead{\colhead{Component}             &
	\colhead{Optical thickness}       &       
           \colhead{$\delta\omega$}      &       
           \colhead{Fringe origin}      &       
           \colhead{Peak-to-peak amplitude }       \\       
           \colhead{}   &
	\colhead{(cm)}       &       
           \colhead{(cm$^{-1}$)}      &       
           \colhead{}      &       
           \colhead{(\%)}       \\       
  	 }
\startdata
 1  &  $0.35 \pm 0.02$  &  $2.8 (\pm 0.2)$  & Detector Substrate &  $\le 40$\\
 2  &  $2.7 \pm 0.2$      &  $0.37 (\pm 0.03)$    &  Dichroic   & $\le 10$  \\
 3  &  0.01 to 0.1          & 10 to 100  &  Fringe beating &  $\le 10$   \\

\enddata
\end{deluxetable}

\begin{deluxetable}{lccc}
\tabletypesize{\footnotesize}
\tablecolumns{5}
\tablewidth{0pt}
\tablecaption{Mean photon conversion efficiency (electron/photon) for all MRS bands. The PCE was averaged over all reconstructed spectral cube sample elements and all wavelengths.}
\tablehead{\colhead{}             &
	\colhead{}       &       
           \colhead{Band }      &       
           \colhead{ }       \\       
           \colhead{}   &
	\colhead{A}       &       
           \colhead{B}      &       
           \colhead{C}       \\       
  	 }
\startdata
Channel 1  &  0.117  &  0.117  & 0.138  \\
Channel 2  &  0.114   & 0.130  & 0.134  \\
Channel 3  &  0.115   & 0.108  & 0.114  \\
Channel 4  & 0.028  & 0.020  & 0.011  \\
\enddata
\end{deluxetable}

\clearpage

\begin{figure}[htbp]
\centerline{\includegraphics[width=7.0in]{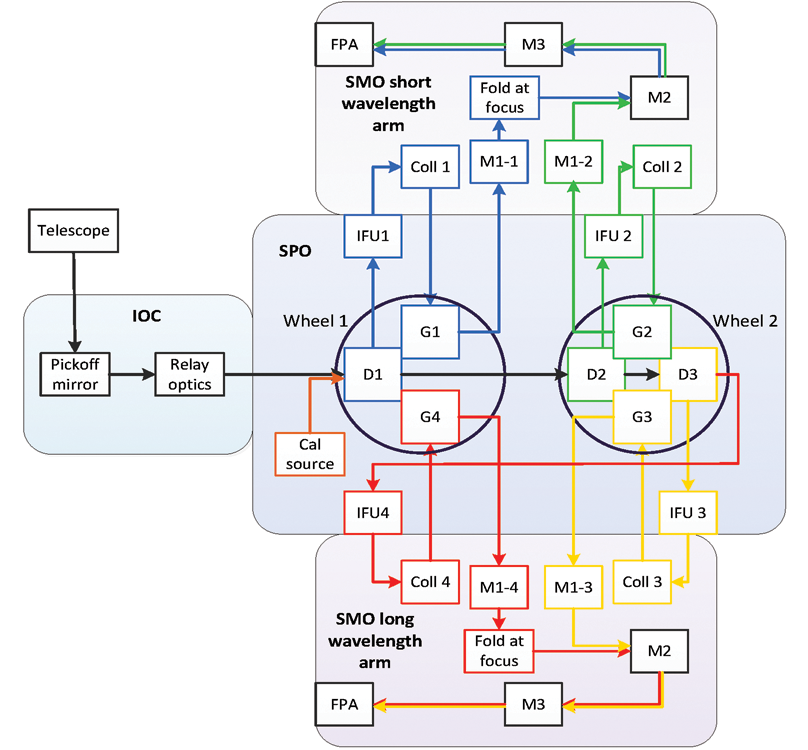}}
%\centerline{\includegraphics[width=7.0in]{f1.eps}}
\caption{Block diagram of the MRS, showing the main optical functions.  A detailed description can be found in the text. }
\label{fig1}
\end{figure}

\clearpage

\begin{figure}[htbp]
\centerline{\includegraphics[width=5.0in]{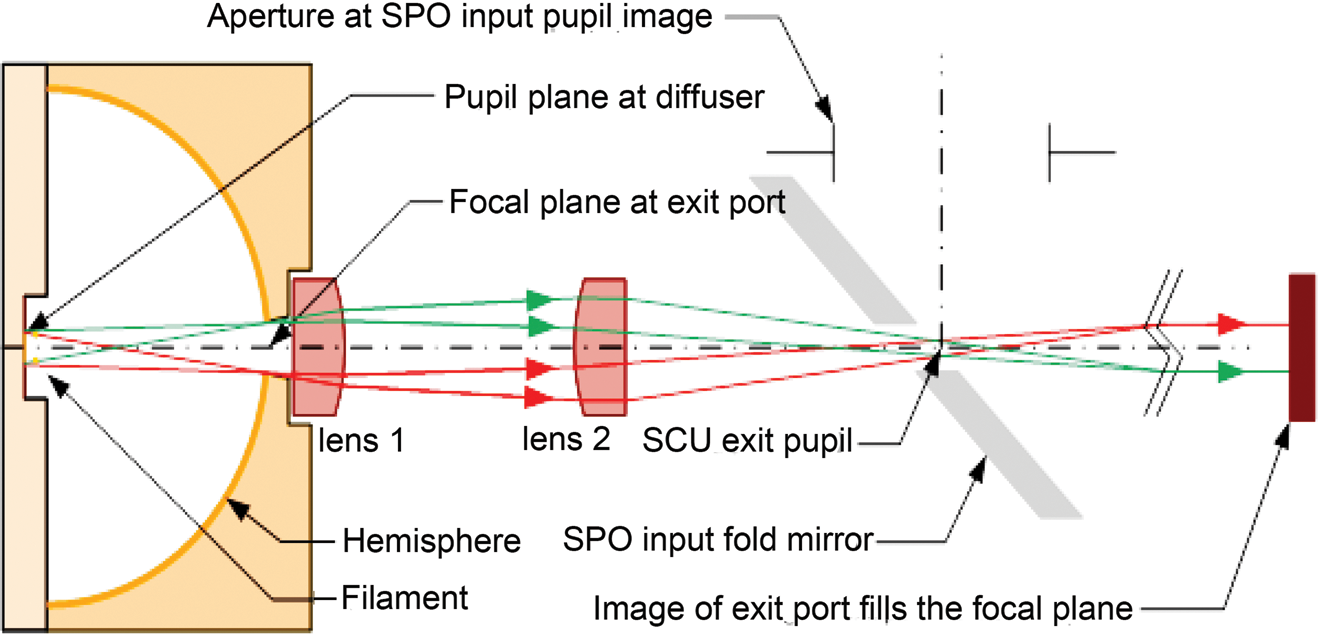}}
%\centerline{\includegraphics[width=5.0in]{f2.eps}}
\caption{Optical Schematic (not to scale) of the Spectrometer Calibration Unit (SCU).  Calibration light is injected via the obscuration at the centre of the JWST pupil where it is imaged at the input fold mirror of the Spectrometer Pre-Optics (SPO). }
\label{fig2}
\end{figure}

\clearpage

\begin{figure}[htbp]
\centerline{\includegraphics[width=7.0in]{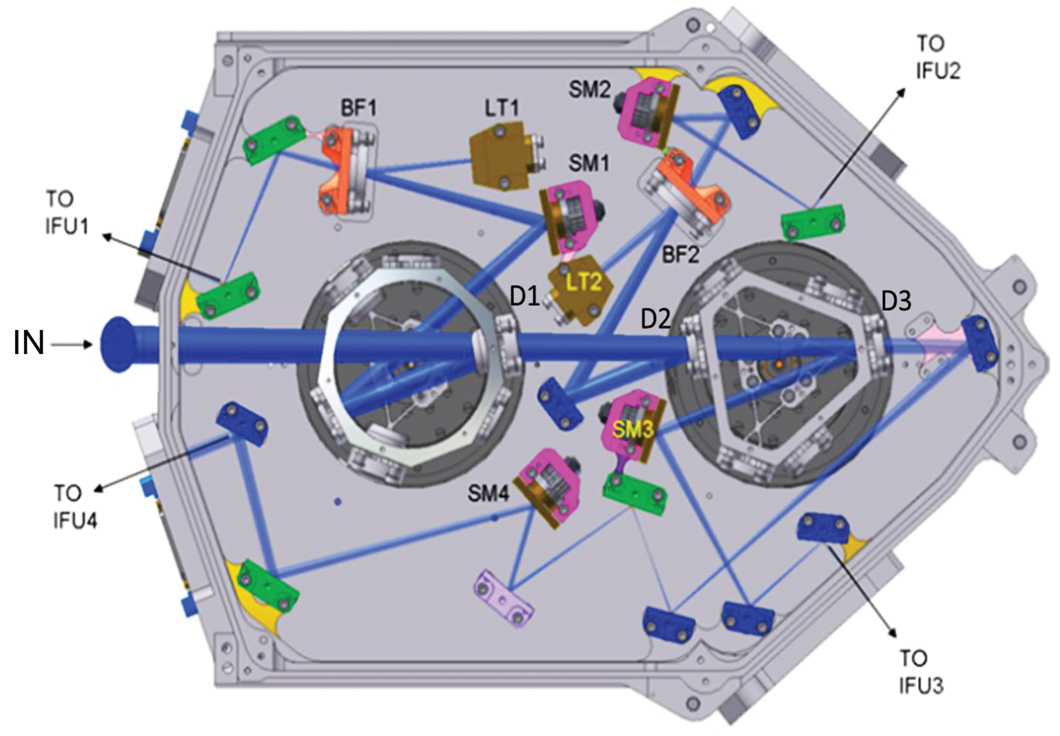}}
%\centerline{\includegraphics[width=7.0in]{f3.eps}}
\caption{ Layout of the Dichroics and Fold Mirrors for all channels in the Spectrometer Pre-Optics (SPO).  The position of the input pupil and fold mirror are labelled ‘IN’. The locations of blocking filters (BF), light traps (LT), powered mirrors (SM) and dichroics (D) are shown. After
the light has been divided into the appropriate spectral ranges, it is output to the integral 
field units (IFU) for the four spectrometer channels.}
\label{fig3}
\end{figure}

%\clearpage

%\begin{figure}[htbp]
%\centerline{\includegraphics[width=5.0in]{fig4.png}}
%\centerline{\includegraphics[width=5.0in]{f4.eps}}
%\caption{Nominal fields of view and sampling of the 4 IFUs. The JWST PSF at the shortest wavelength wavelength in each channel is overlaid by the sampling elements. }
%\label{fig4}
%\end{figure}

\clearpage

\begin{figure}[htbp]
\centerline{\includegraphics[width=5.0in]{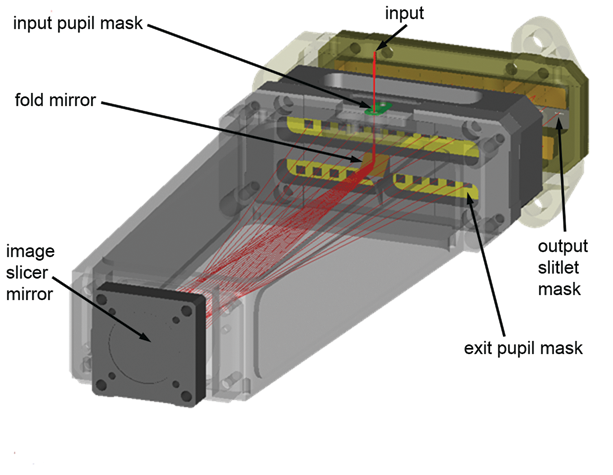}}
%\centerline{\includegraphics[width=5.0in]{f5.eps}}
\caption{ Three Dimensional View of the Channel 3 IFU.  }
\label{fig5}
\end{figure}

\clearpage

\begin{figure}[htbp]
\centerline{\includegraphics[width=5.0in]{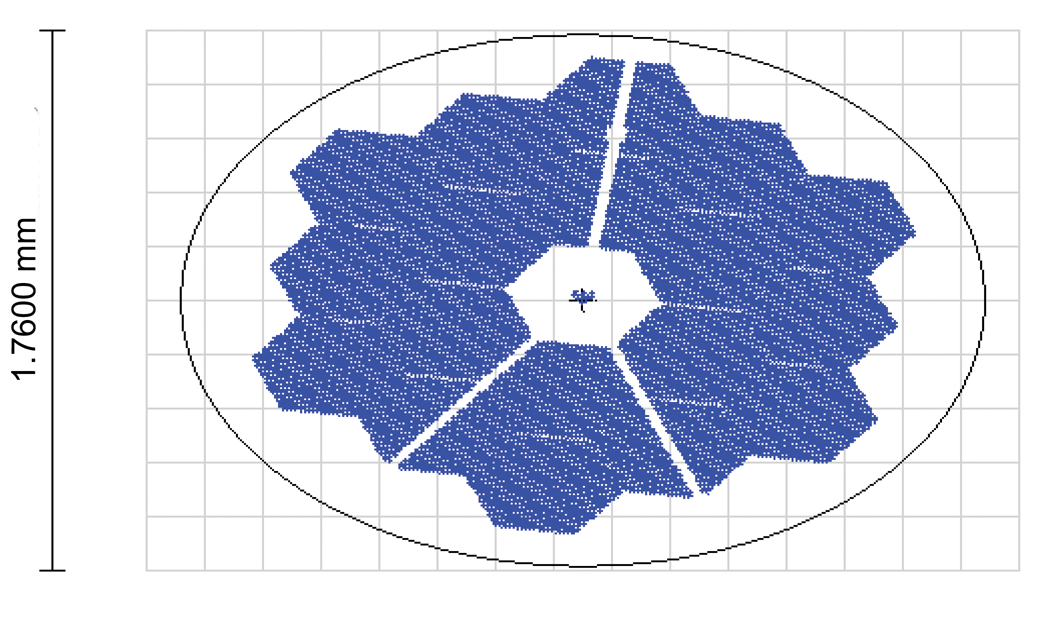}}
%\centerline{\includegraphics[width=5.0in]{f6.eps}}
\caption{ Rotated and anamorphically magnified pupil image at the entrance to the IFU of Channel 3, plotted on a square grid.}
\label{fig6}
\end{figure}

\clearpage

\begin{figure}[htbp]
\centerline{\includegraphics[width=5.0in]{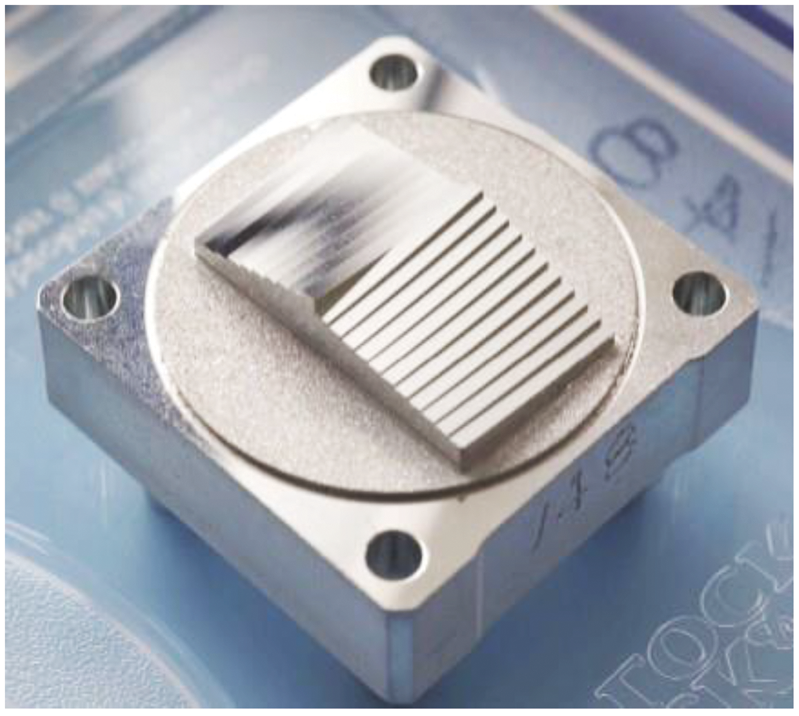}}
%\centerline{\includegraphics[width=5.0in]{f7.eps}}
\caption{Photograph of the Channel 1 Image Slicer Mirror (Courtesy of Cranfield University). 
The slices are 1mm wide and 12 mm long.}
\label{fig7}
\end{figure}

\clearpage

\begin{figure}[htbp]
\centerline{\includegraphics[width=5.0in]{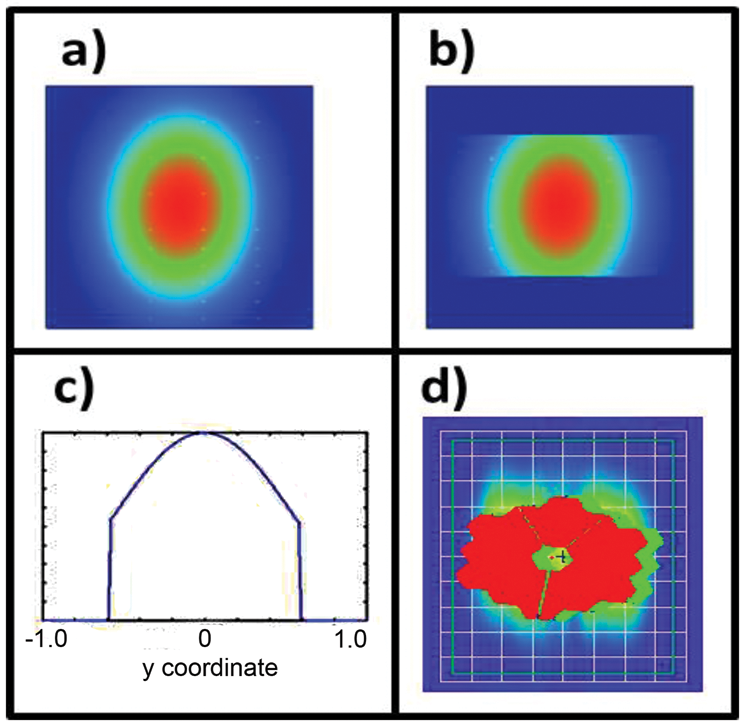}}
%\centerline{\includegraphics[width=5.0in]{f8.eps}}
\caption{(a) and (b), Input and truncated images on the imager slicer mirror.  (c), Cross-section through the peak of the truncated image.  (d),  Diffraction broadened pupil images (see text for details). }
\label{fig8}
\end{figure}

\clearpage

\begin{figure}[htbp]
\centerline{\includegraphics[width=7.0in]{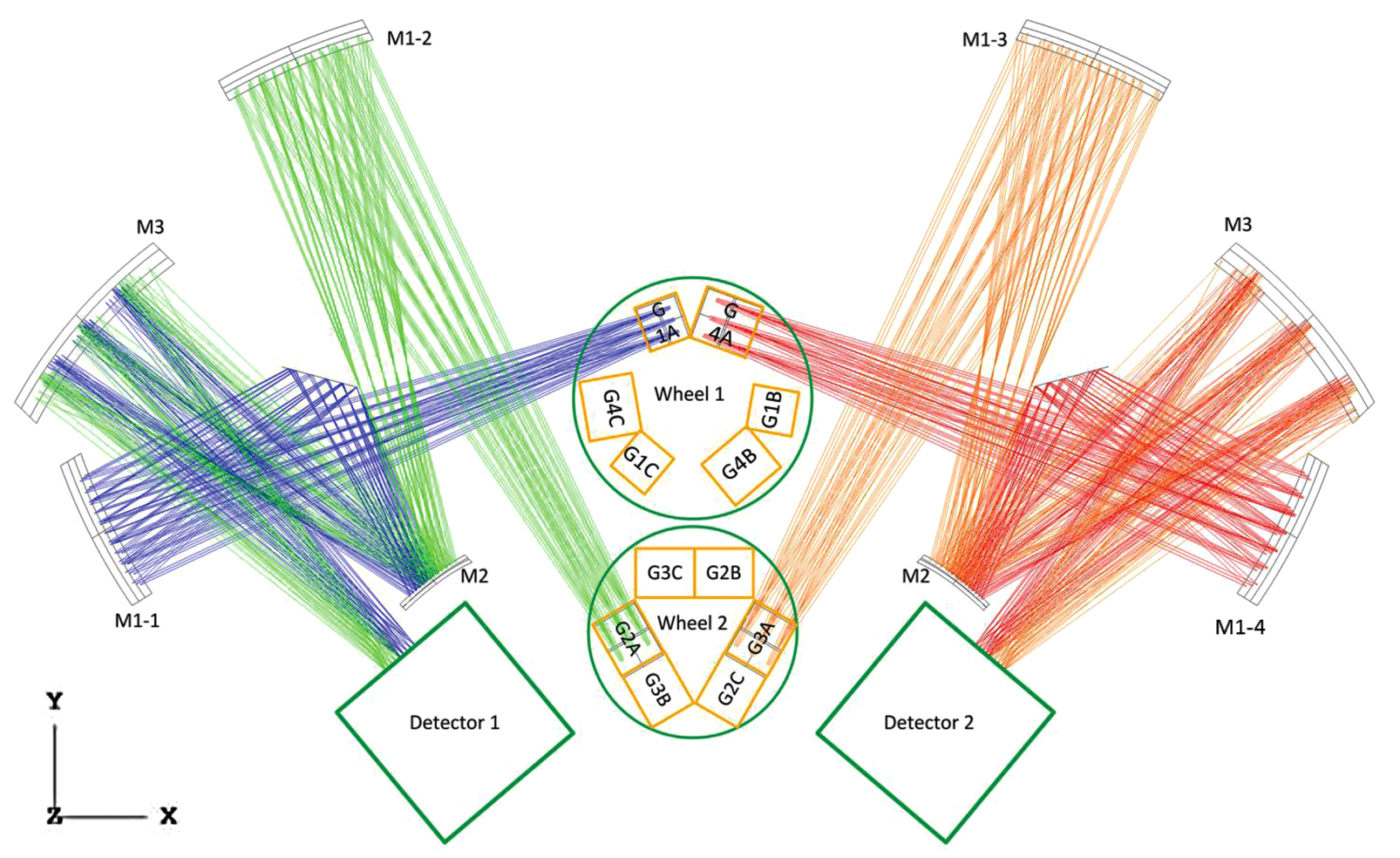}}
%\centerline{\includegraphics[width=7.0in]{f9.eps}}
\caption{SMO cross-section of the camera symmetry plane in the spatial direction.  Light arrives at the gratings (G1A, G2A, G3A, G4A) from the IFU output collimating mirrors.}
\label{fig9}
\end{figure}

\clearpage

\begin{figure}[htbp]
\centerline{\includegraphics[width=5.0in]{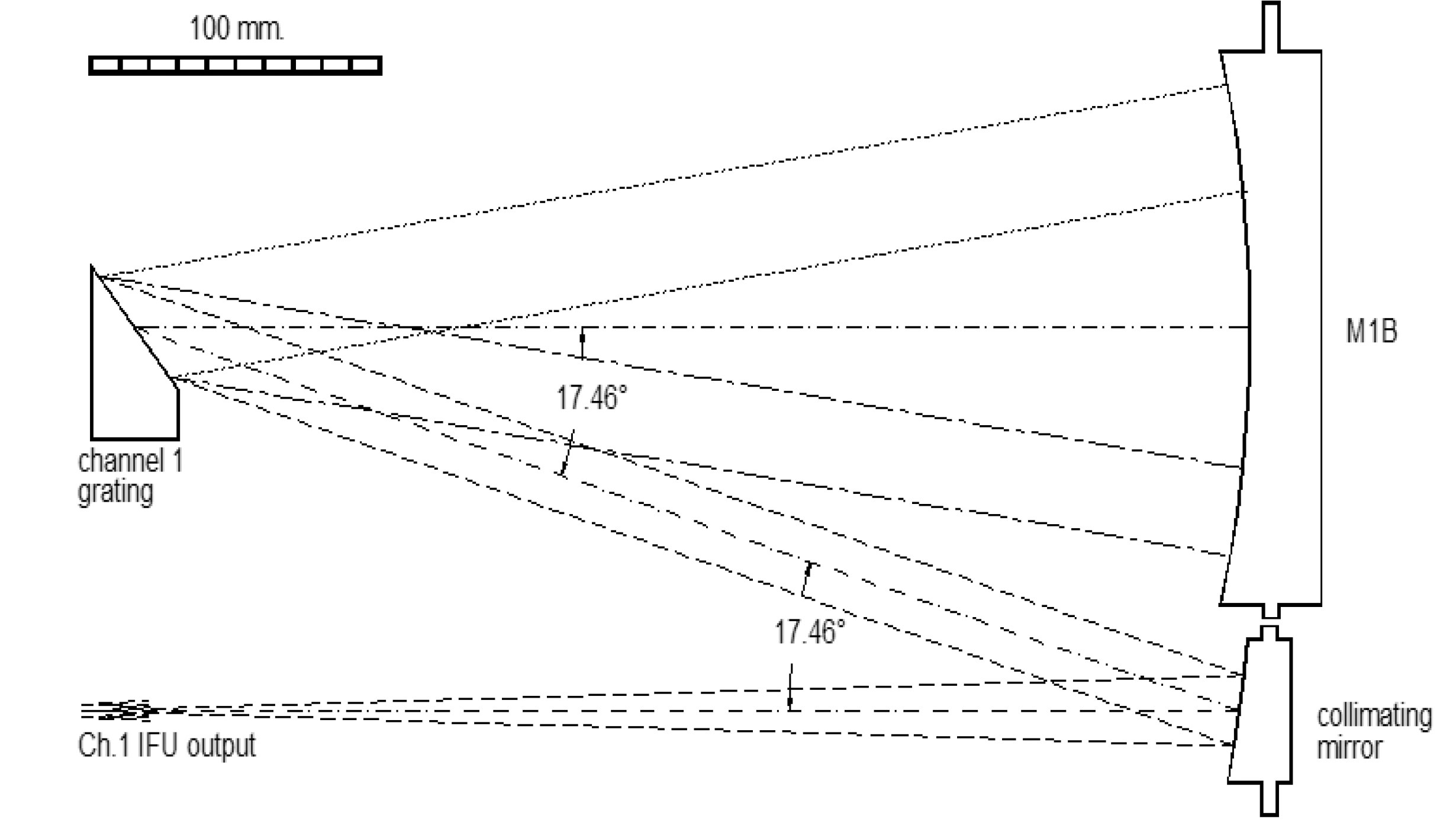}}
%\centerline{\includegraphics[width=5.0in]{f10.eps}}
\caption{Cross-section along the dispersion direction of the collimator, grating and camera mirror M1 of the SMO. }
\label{fig10}
\end{figure}

\clearpage

\begin{figure}[htbp]
\centerline{\includegraphics[width=5.0in]{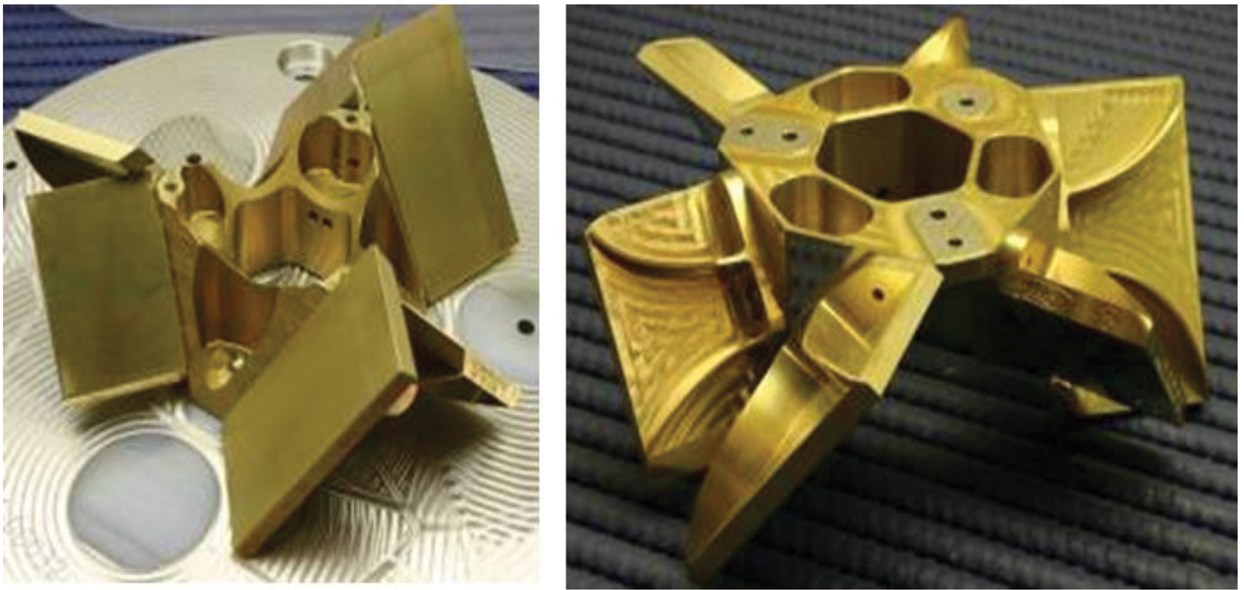}}
%\centerline{\includegraphics[width=5.0in]{f11.eps}}
\caption{Grating Wheel Assembly (Wheel 1). }
\label{fig11}
\end{figure}

\clearpage

\begin{figure}[htbp]
\centerline{\includegraphics[width=7.0in]{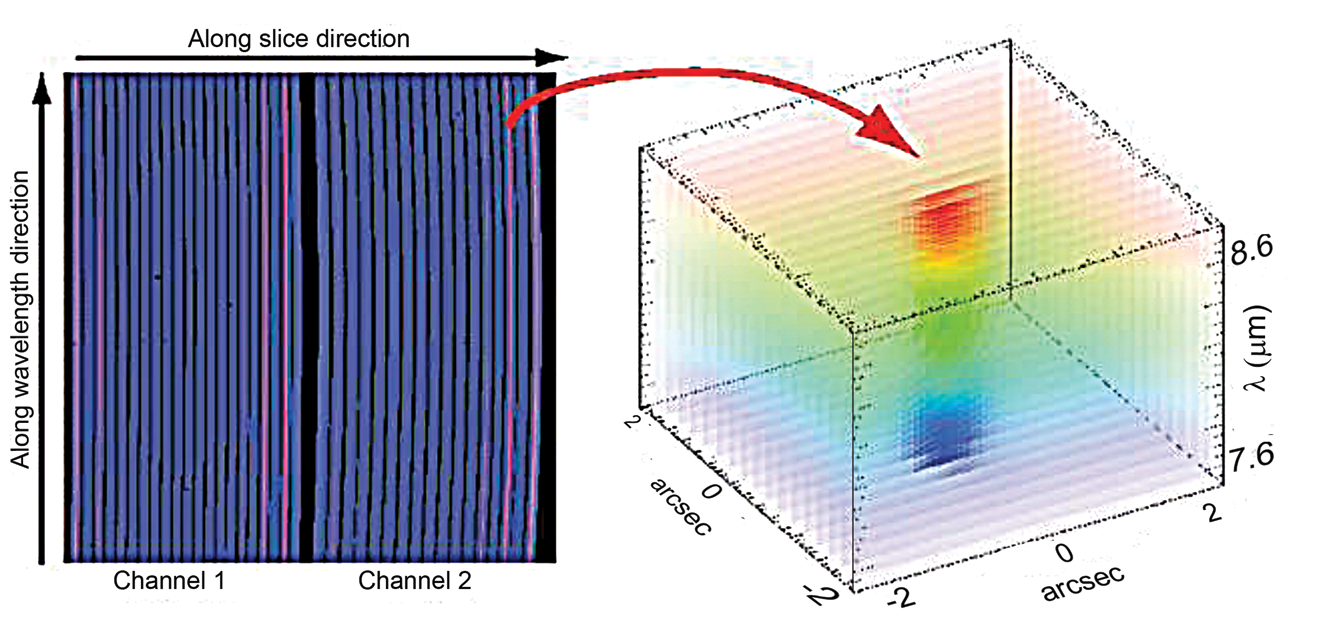}}
%\centerline{\includegraphics[width=7.0in]{f12.eps}}
\caption{Illustration of the reconstruction process: The detector image for two channels is reconstructed into a 3D cube with two spatial dimensions ($\alpha$ along the slice and $\beta$ across the slice) and one spectral dimension ($\lambda$).}
\label{fig12}
\end{figure}

\clearpage

\begin{figure}[htbp]
\centerline{\includegraphics[width=5.0in]{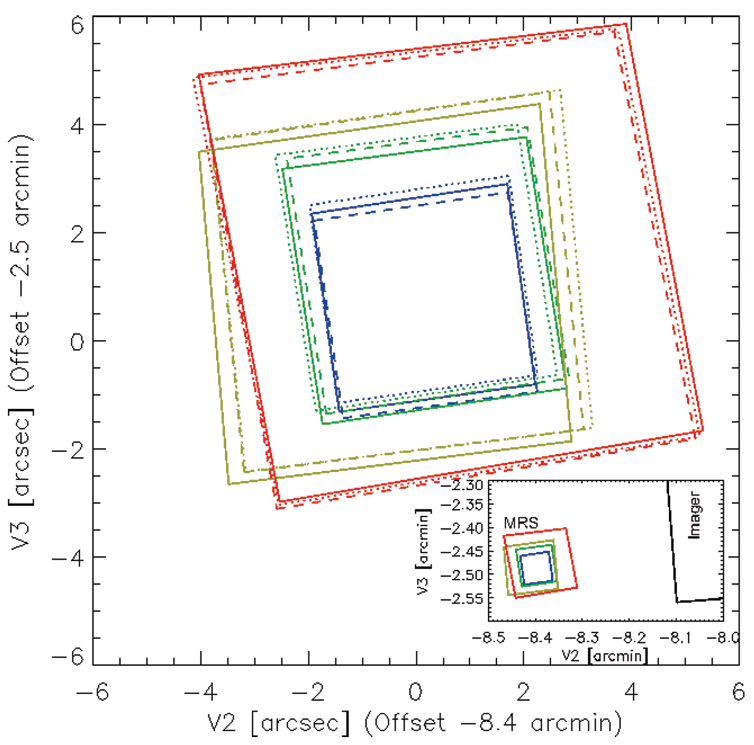}}
%\centerline{\includegraphics[width=5.0in]{f13.eps}}
\caption{Transformed MRS fields in the JWST coordinate frame (V2, V3). The field borders are drawn as solid lines (Sub-band A), dashes (Sub-band B) and dots (Sub-band C) for channel 1 (blue), 2 (green), 3 (yellow) and 4 (red).  The inset shows the position of the MRS FOV relative to the imager field.}
\label{fig13}
\end{figure}

\clearpage

\begin{figure}[htbp]
\centerline{\includegraphics[width=5.0in]{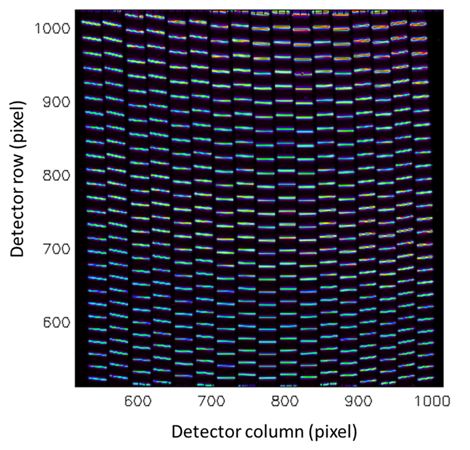}}
%\centerline{\includegraphics[width=5.0in]{f14.eps}}
\caption{A typical wavelength calibration observation.  The long wavelength half of Sub-band 2A is shown when illuminated by an extended continuum source modulated by a vacuum spaced etalon with a free spectral range (fringe spacing) of 3.1 cm$^{-1}$ at ambient. }
\label{fig14}
\end{figure}

\clearpage

\begin{figure}[htbp]
\centerline{\includegraphics[width=7.0in]{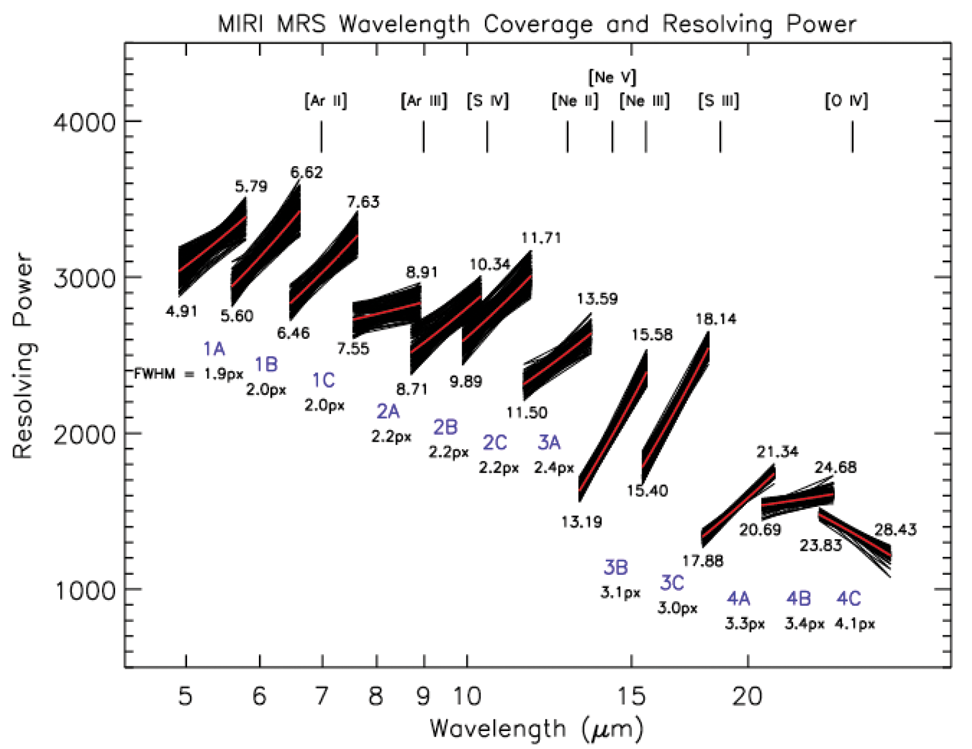}}
%\centerline{\includegraphics[width=7.0in]{f15.eps}}
\caption{The spectral resolving power of the MRS.  The black regions represent variations in the resolving power across the FOV for each sub-band, and the red lines are spatially averaged values. The wavelength ranges of each sub-band are indicated, as well as some scientifically important spectral lines.  The mean width (FWHM) of the spectral resolution element in pixels is also provided for each sub-band.}
\label{fig15}
\end{figure}

\clearpage

\begin{figure}[htbp]
\centerline{\includegraphics[width=5.0in]{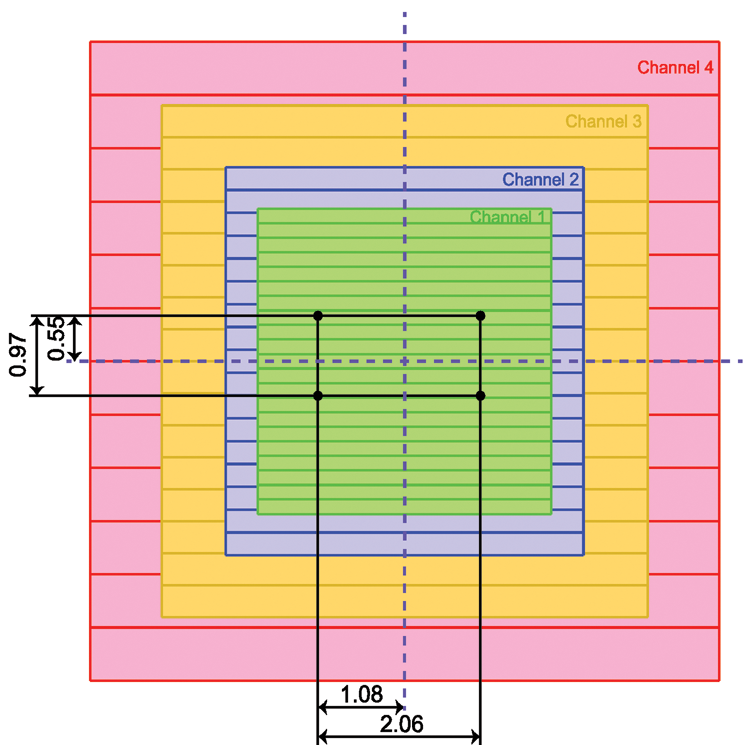}}
%\centerline{\includegraphics[width=5.0in]{f16.eps}}
\caption{Coordinates in arcsec of the four nominal dither positions relative to the centre of the MRS field (dashed axes). }
\label{fig16}
\end{figure}

\clearpage

\begin{figure}[htbp]
\centerline{\includegraphics[width=5.0in]{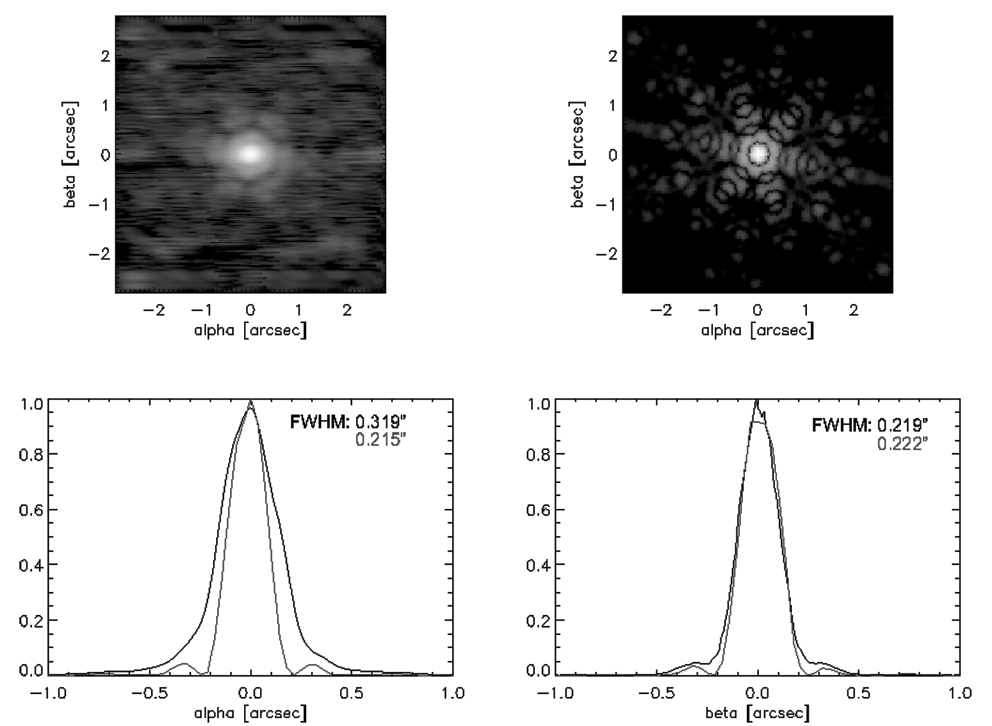}}
%\centerline{\includegraphics[width=5.0in]{f17.eps}}
\caption{Measured PSF at 5.6 $\mu$m (top left) compared with the FFT of the JWST pupil (top right) and normalized peak profiles of the measured (black) and modelled (red) for the along-slice (bottom left) and across-slice direction (bottom right). The X-axis is multiplied by two for the lower figures.}
\label{fig17}
\end{figure}

\clearpage

\begin{figure}[htbp]
\centerline{\includegraphics[width=5.0in]{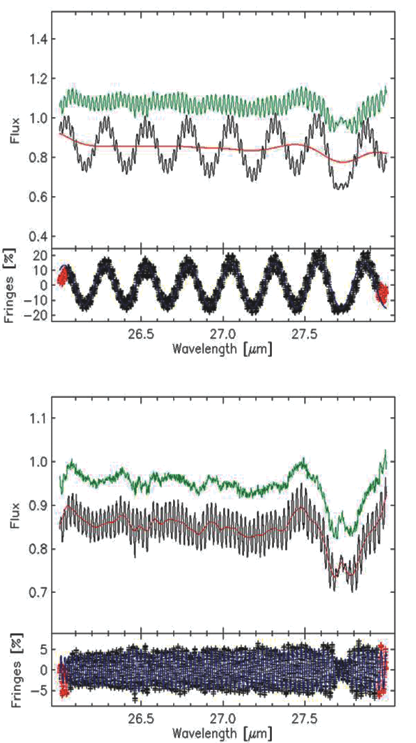}}
%\centerline{\includegraphics[width=5.0in, height=6.0in]{f18.eps}}
\caption{Examples of two main observed MRS fringe components. In the top panels of each plot the input data is shown in black with da fitted continuum in red. The spectrum (offset for clarity) after fringe removal is shown in green. In the lower panels the normalized spectrum and fitted fringes are shown.  Top: the main component \#1 in Channel 4 (after de-fringing, fringe component \#2 remains); bottom: the high frequency fringe component \#2.}
\label{fig18}
\end{figure}

\clearpage

\begin{figure}[htbp]
\centerline{\includegraphics[width=5.0in]{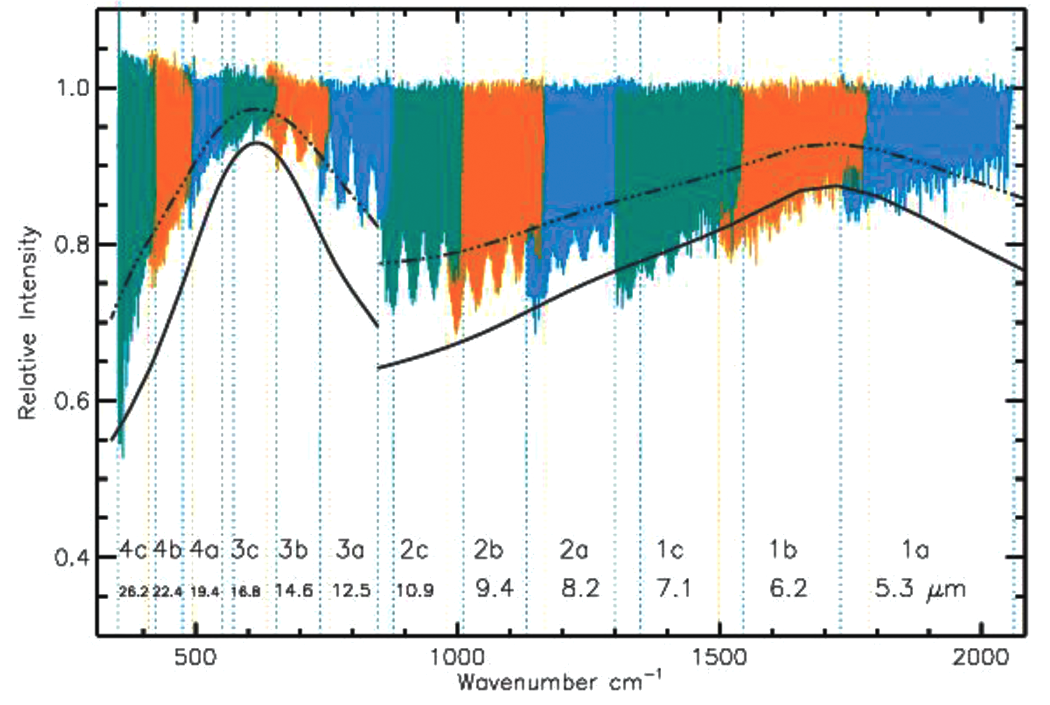}}
%\centerline{\includegraphics[width=5.0in]{f20.eps}}
\caption{The combined fringes for each channel and sub-band. The data for each neighbouring band are plotted in a different colour for clarity. The main and high frequency fringe components are not resolved on this scale.  In Channels 2 and 3 an example of a low frequency beating fringe can clearly be seen.  Note: the spectrum is normalized to the peaks of the main fringe amplitude. See text for a description of the solid and dashed-dotted lines.}
\label{fig19}
\end{figure}

\clearpage

\begin{figure}[htbp]
\centerline{\includegraphics[width=5.0in]{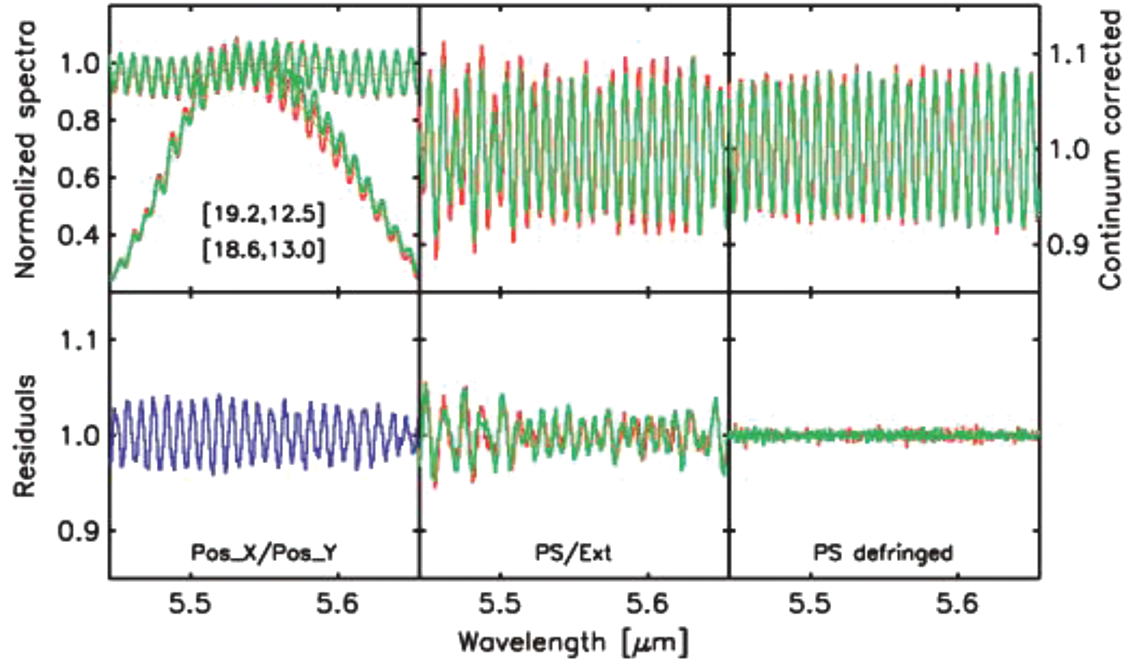}}
%\centerline{\includegraphics[width=5.0in]{f19.eps}}
\caption{ Illustration of the fringe source dependence for point and extended blackbody source spectra at two positions (red and green spectra) as printed in the top-left panel.  Top from left to right; peak normalized point and extended spectra; normalized point source spectra; normalized extended blackbody spectra. Bottom from left to right; both point source spectra divided; point source spectra divided by extended source spectra (akin to direct flat-field correction); both normalized point source spectra de-fringed using a sine fitting technique.}
\label{fig20}
\end{figure}

%\clearpage

%\begin{figure}[htbp]
%\centerline{\includegraphics[width=5.0in]{fig21.png}}
%\centerline{\includegraphics[width=5.0in]{f21.eps}}
%\caption{A sample of MRS data showing the effects of the fringes (left panel) and after the fringe correction (right).}
%\end{figure}

\clearpage

\begin{figure}[htbp]
\centerline{\includegraphics[width=5.0in]{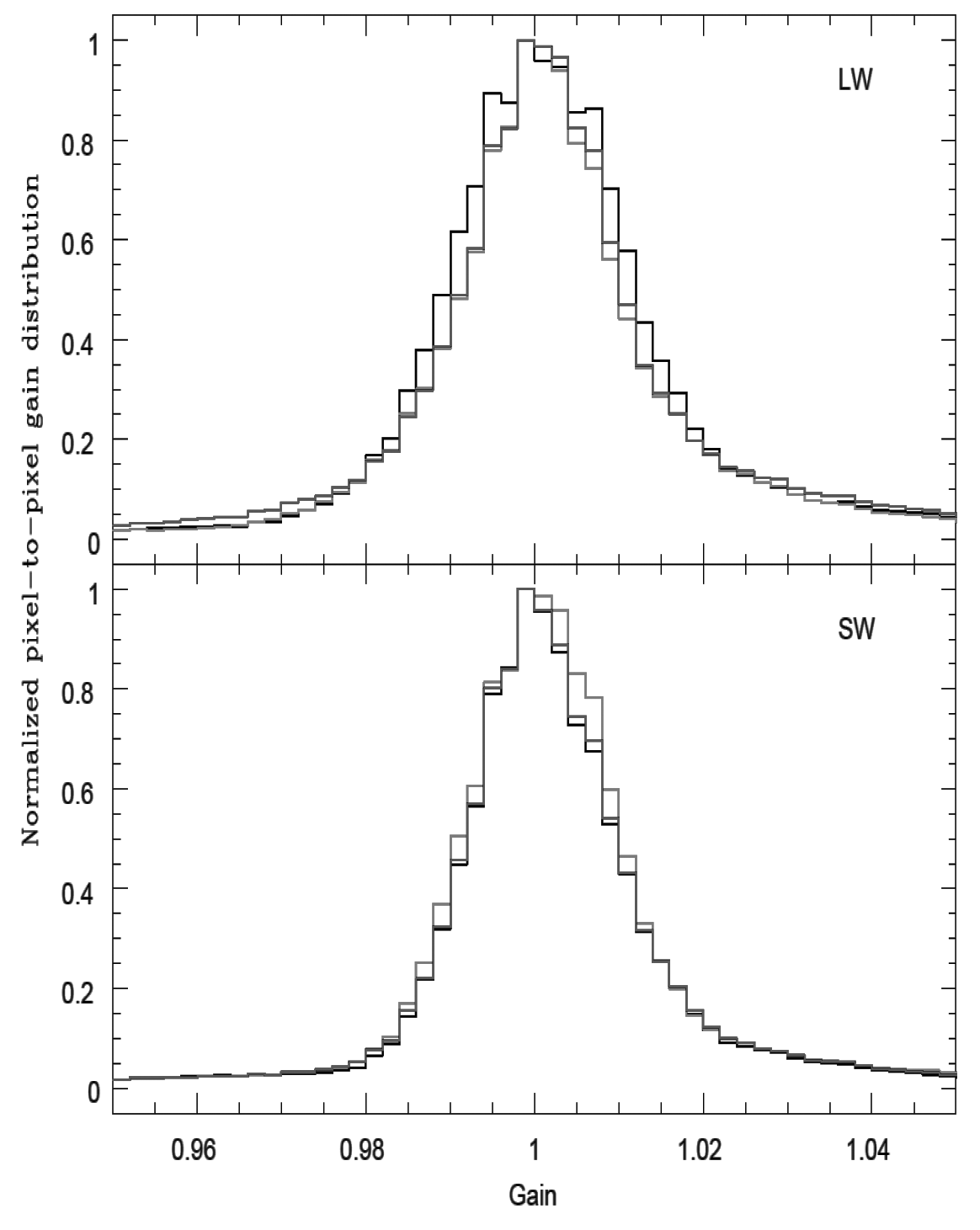}}
%\centerline{\includegraphics[width=5.0in, height=4.5in]{f22.eps}}
\caption{Normalized pixel-to-pixel gain distributions derived from 3 
observations for SW and LW detectors (black, red and blue). The FWHMs are 
1.91 {\%} and 1.99 {\%} for the SW and LW channels, respectively.}
\end{figure}

\clearpage

\begin{figure}[htbp]
\centerline{\includegraphics[width=5.0in]{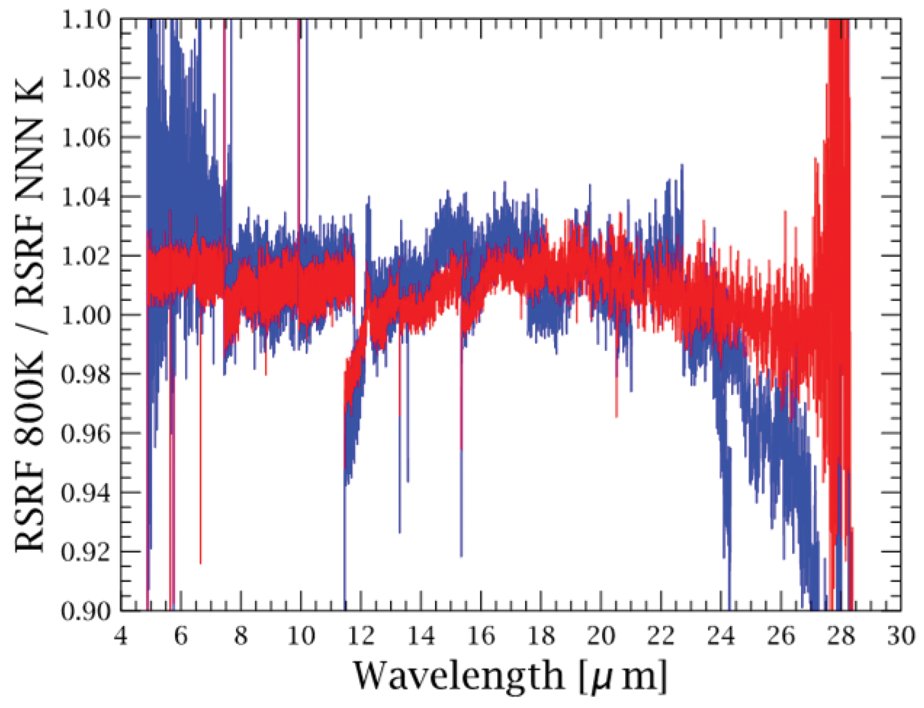}}
%\centerline{\includegraphics[width=5.0in, height=3.5in]{f23.eps}}
\caption{Comparison of the Relative Spectral Response function calculated using observations of the MTS black body at 400 K (blue) and at 600 K (red) with the RSRF at 800 K.}
\label{fig23}
\end{figure}

\clearpage

\begin{figure}[htbp]
\centerline{\includegraphics[width=5.0in]{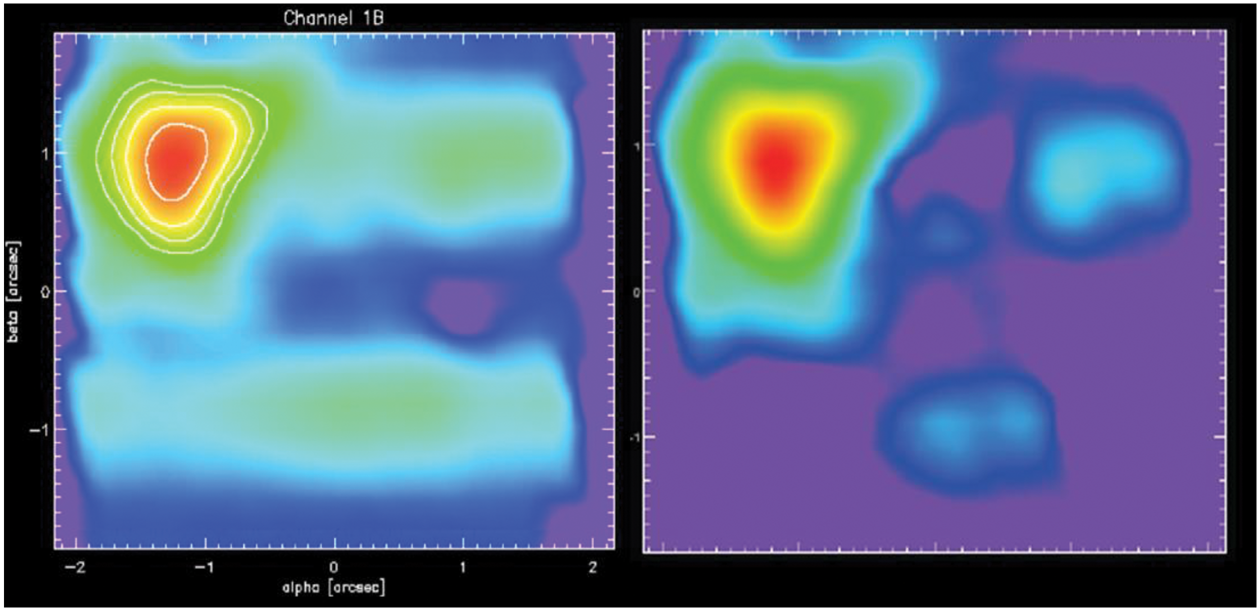}}
%\centerline{\includegraphics[width=5.0in]{f24.eps}}
\caption{Straylight in Channel 1B.  Left: The reconstructed image of a bright, moderately compact source is plotted on a logarithmic scale with contours at 50, 20, 10 and 5 \% of the peak signal level.  Right: The image after stray-light correction using a prototype algorithm.}
\label{fig24}
\end{figure}

\clearpage

\begin{figure}[htbp]
\centerline{\includegraphics[width=5.0in]{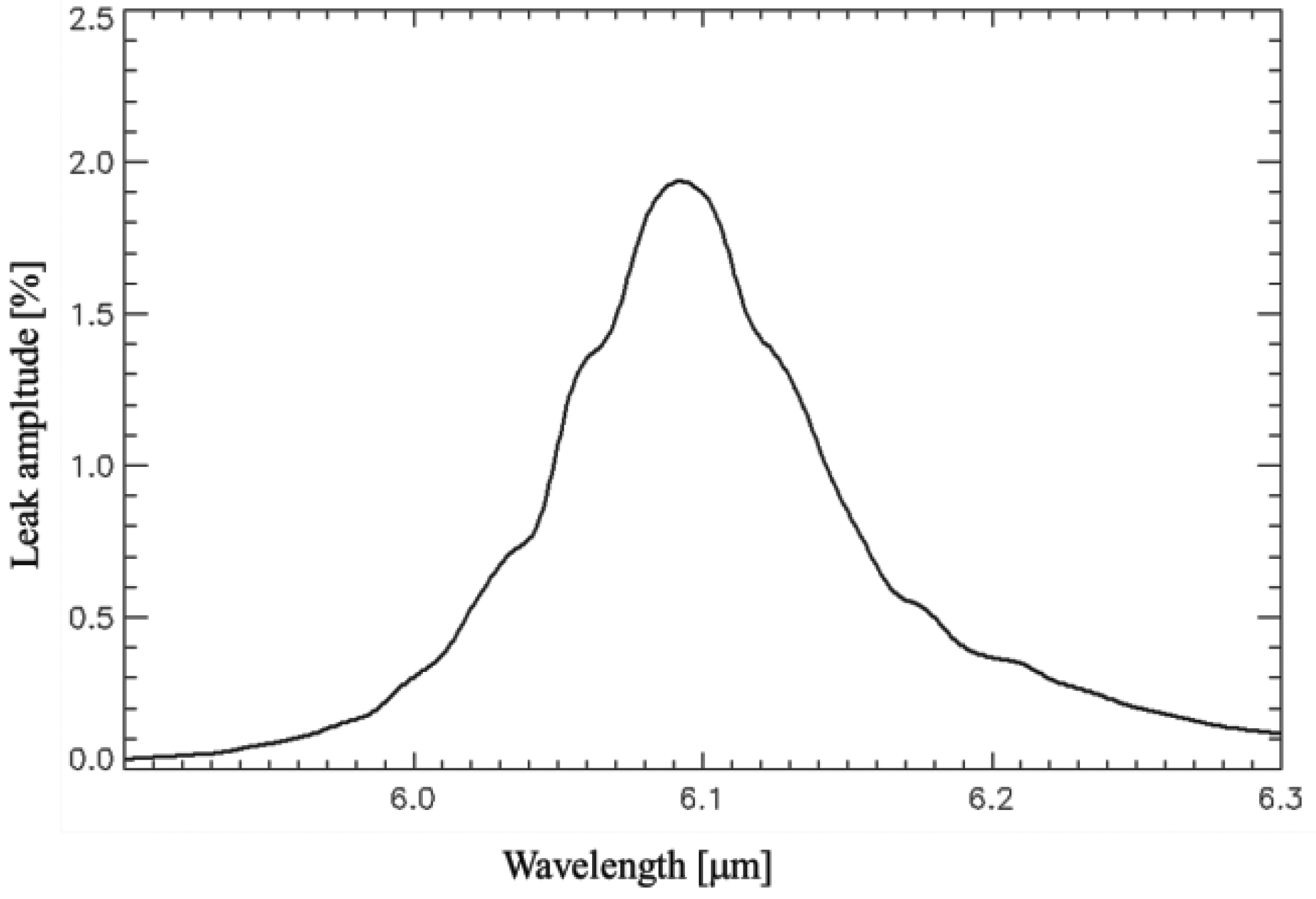}}
%\centerline{\includegraphics[width=5.0in]{f25.eps}}
\caption{MRS Channel 3A spectral leak profile.  This is the fraction of the source spectrum at $\lambda$ = 6.1 $\mu$m which is added to the nominal 12.2 $\mu$m spectrum.}
\label{fig25}
\end{figure}

\end{document}